\documentclass[letterpaper,aps,pra,10pt,superscriptaddress,showpacs,floats,twocolumn,nofootinbib,reprint]{revtex4-1}
\usepackage[latin1]{inputenc}
\usepackage[T1]{fontenc}
\usepackage{amsmath,amssymb,mathbbol}

\usepackage{graphicx,color}
\usepackage{hyperref}

\newcommand{\titlestr}{Nanoplasmonic near-field synthesis}

\hypersetup{colorlinks,linkcolor=blue,citecolor=blue,urlcolor=blue,hypertexnames=true,%
pdfauthor={Johannes Feist, M. T. Homer Reid, Matthias F. Kling},pdftitle={\titlestr}}

\begin{document}

\newcommand{\etal}{\emph{et~al.}\ }
\newcommand{\ie}{i.e., }
\newcommand{\eg}{e.g., }
\newcommand{\cf}{cf.\ }

\newcommand{\eV}{\,\text{eV}}
\newcommand{\nm}{\,\text{nm}}
\newcommand{\Wcm}{\,\text{W}/\text{cm}^2}
\newcommand{\fs}{\,\text{fs}}

\renewcommand{\equationautorefname}{Eq.}
\renewcommand{\figureautorefname}{Fig.}
\renewcommand{\sectionautorefname}{Sec.}

\title{\titlestr}

\author{Johannes~Feist}
\email{johannes.feist@uam.es}
\affiliation{ITAMP, Harvard-Smithsonian Center for Astrophysics, Cambridge, Massachusetts 02138, USA}
\affiliation{Departamento de F\'isica Te\'orica de la Materia Condensada, Universidad Aut\'onoma de Madrid, E-28049 Madrid, Spain}

\author{M.~T.~Homer~Reid}
\affiliation{Department of Mathematics, Massachusetts Institute Of Technology, Cambridge, Massachusetts 02139, USA}

\author{Matthias~F.~Kling}
\affiliation{Max Planck Institute of Quantum Optics, D-85748 Garching, Germany}
\affiliation{J.R. Macdonald Laboratory, Kansas-State University, Manhattan, Kansas 66506, USA}

\date{\today}

\begin{abstract}
The temporal response of resonances in nanoplasmonic structures typically converts an incoming few-cycle field into a much longer near-field 
at the spot where non-linear physical phenomena including electron emission, recollision and high-harmonic generation can take place. We show that for practically useful structures pulse shaping of the incoming pulse can be used to synthesize the plasmon-enhanced field and enable single-cycle driven nonlinear physical phenomena. Our method is demonstrated for the generation of an isolated attosecond pulse by plasmon-enhanced high harmonic generation. We furthermore show that optimal control techniques can be used even if the response of the plasmonic structure is not known \emph{a priori}.
\end{abstract}

\maketitle

\section{Introduction}
In the last decade, there has been rapid progress in the production of ultrashort light pulses with controlled waveforms and durations down to below $100$~attoseconds~\cite{KraIva2009,GouSchHof2008,ZhaZhaChi2012}. These light sources enable the control and tracing of electron dynamics in atoms ~\cite{SchFieKar2010,GouLohWir2010,KluDahGis2011}, molecules~\cite{KliSieVer2006,SanKelPer2010}, and solids~\cite{ApoDomPau2004,CavMueUph2007} on their natural time scale. The realization of a similar level of control of the electron motion in nanocircuits has the potential to revolutionize modern electronics~\cite{KraIva2009}. Light-wave-controlled nanocircuits (light-wave nanoelectronics) may reach petahertz operation frequencies and might remove the bottleneck in conventional communication technology by enabling all-optical information processing and communication.
The key to light-wave nanoelectronics is the control of electron dynamics in nanostructured materials on sub-cycle time scales. Progress has very recently been made in the control of electron dynamics in nanoparticles~\cite{ZheFenPle2011}, nanotips~\cite{KruSchHom2011}, and nanojunctions~\cite{SchPaaKar2012} with carrier-envelope phase (CEP) stabilized few-cycle pulses. The control of few-cycle waveforms by the CEP, however, gives only a very limited degree of control over the electron dynamics, which can be significantly improved with ultrabroadband light-wave synthesis permitting to sculpt the electric field of a laser pulse with attosecond precision~\cite{WirHasGrg2011}. Despite its importance for light-wave nanoelectronics, this approach has not yet been implemented for the shaping of plasmonic near-fields.

The application of few-cycle pulses to a resonant nanostructure typically results not only in the desired field-enhancement effect, which may be utilized for nonlinear applications, but also in a temporally longer near-field evolution~\cite{SteSueYan2011}. It is desirable, however, to reach single-cycle near-field profiles for many applications, such as a well-controlled asymmetry in the direction of electronic currents, the confinement of electron emission and acceleration to a single cycle, and the generation of isolated attosecond extreme ultraviolet (XUV) pulses. Furthermore, if CEP control of strong-field processes that are enabled by the near-fields is required, the amplitude of those CEP effects typically scale inverse exponentially with the near-field duration, and their relevance is thus limited to the few-cycle regime~\cite{RouEsr2007}.

In this article, we present an approach employing pulse-shaping techniques to form plasmonic near-field transients which enable nonlinear phenomena that are induced by just a single cycle of the near-field. While this approach is general and not limited to a certain nonlinear process, we will discuss its implementation for the generation of isolated attosecond pulses via plasmon-assisted high-harmonic generation (HHG)~\cite{KimJinKim2008,ParKimCho2011,SteSueYan2011,HusKelHer2011,CiaBieQui2012}. Here, the strong field enhancement obtained from plasmon resonances in metallic nanostructures is used to locally enhance the electric field strength to the levels required for HHG. As the local intensity can be enhanced by more than four orders of magnitude, this drastically lowers the required driving laser intensity and enables HHG with repetition rates in the MHz range.\footnote{It should be noted that the experimental studies in~\cite{KimJinKim2008} were recently challenged by Sivis \textit{et al}., who only observed XUV fluorescence from bow-tie nanoantennas for similar conditions~\cite{SivDuwAbe2012}. Park \textit{et al.}~recently provided data aimed at supporting their initial claim~\cite{ParChoLee2013}. Further studies will, however, be required to clarify the importance of coherent versus incoherent XUV emission from such nanostructures. Since HHG is only used as an example application for nanoplasmonic near-field synthesis in our studies, we neglect incoherent processes.} However, due to the temporal distortion of the near field, current approaches do \emph{not} produce isolated attosecond pulses, which are a critical ingredient for attosecond spectroscopy.

As an example, we study one of the structures investigated in Ref.~\cite{SteSueYan2011}: two gold ellipsoids with major and minor axes of $100$ and $16.7\nm$, respectively. The major axes of both ellipsoids are aligned along the same axis, with a gap of $5\nm$ between them. This geometry creates a local ``hot spot'' between the ellipsoids, where the field enhancement is expected to be maximal.
In order to support the generation of isolated attosecond pulses, the high-harmonic generation process has to be \emph{gated} so that it only occurs within a short window of time. Husakou \etal have recently shown that one common approach, \emph{polarization gating}~\cite{CorBurIva1994,ShaGhiCha2005,SanBenCal2006}, can be transferred to the nanoscale for plasmonic structures that replicate the polarization properties of the incoming pulse~\cite{HusKelHer2011}.
We here propose a more general approach using pulse shaping such that the plasmon-enhanced near field achieves \emph{amplitude gating}~\cite{Christov97,BraKra2000}, where the generating near field only becomes strong enough to generate high-energy harmonics during a short time window. An isolated attosecond pulse can then be generated by spectrally filtering the generated high harmonic radiation. This approach is quite independent of the specific properties of the plasmonic structure. One main finding is that even when the lifetime of the plasmon resonance is much larger than the cycle time of the IR field, the local response can be shaped to allow for the generation of isolated attosecond pulses with low noise, while still exploiting the large plasmonic field enhancement.
We stress that while we focus on amplitude gating, the pulse distortion also prevents straightforward application of other techniques for isolated attosecond pulse generation, such as two-color gating or double optical gating~\cite{MasGilLi2008,FenGilMas2009}. For these methods, the distortion could also be compensated along similar lines as presented in the following.

The paper is organized as follows: In \autoref{sec:method}, we introduce the theoretical methods used. In \autoref{sec:results}, we present our results on the frequency-dependent response of the system and on isolated attosecond pulse generation. Our findings are summarized in \autoref{sec:summary}.

\section{Method}\label{sec:method}
The gold ellipsoids are described in the limit of local linear dielectric response, with a frequency-dependent dielectric function $\varepsilon_r(\omega)$
taken from the experimental (bulk) values in Ref.~\cite{CRCHandbook}. While retardation is fully included, we neglect non-local effects, which can decrease the maximum field enhancement close to very small features \cite{McmGraSch2009,McMGraSch2010,DavGar2011}.
The strongly nonlinear response of the atomic gas used as the HHG medium is small enough that it can be neglected when solving the Maxwell equations. Because of the linearity of the Maxwell equations in this approximation, the plasmon response to an ultrashort (broad-band) incoming pulse can be calculated as the superposition of fixed-frequency components with appropriate amplitudes. The spatiotemporal electric field distribution is then given by
\begin{equation}\label{eq:fourier_synth}
\vec E(\vec r,t) = \sum_n c_n \vec E_n(\vec r) e^{-i\omega_n t} + c.c.\,,
\end{equation}
where $\vec E_n(\vec r)$ is the (complex-valued) spatial response for incoming mode $n$ at frequency $\omega_n$, and $c_n$ are the complex amplitudes determining the temporal shape of the incoming pulse. We choose all incoming modes to be plane waves polarized along the axis connecting the ellipsoids (the $z$ axis) and
propagating along the same orthogonal direction ($x$ axis). Thus, $\vec E_n(\vec r) \approx\hat z e^{i w_n x/c}$ for large negative $x$. Using plane waves is equivalent to assuming that the focus spot size is large compared to the extension of the system. The response $\vec E_n(\vec r)$ for each fixed-frequency component is obtained using the \textsc{Scuff-EM} package~\cite{scuff-em, ReiRodWhi2009}, a free, open-source implementation of the boundary-element method (BEM) of classical electromagnetic scattering~\cite{Har1993}.  The BEM exploits known Maxwell solutions to express the fields inside and outside homogeneous material bodies in terms of effective \textit{surface currents} flowing on the body surfaces. This has the advantage that we need only discretize surfaces, not volumes, yielding a computationally efficient approach.

The frequency components used to construct (synthesize) the pulses are integer multiples of $\omega_0=0.031\eV$ (giving a \emph{frequency comb}).
We use $79$ frequency components $\omega_n = n\,\omega_0$ with $n\in [8,86]$. The period $T=2\pi/\omega_0 \approx 133 \fs$ of the resulting incoming pulse train is long enough to ensure that the response to each pulse is independent. We checked that using half the period leaves the results essentially unchanged. The periodicity thus does not influence our conclusions.

\section{Results}\label{sec:results}
\subsection{Frequency-dependent response}
\begin{figure}
  \includegraphics[width=\linewidth]{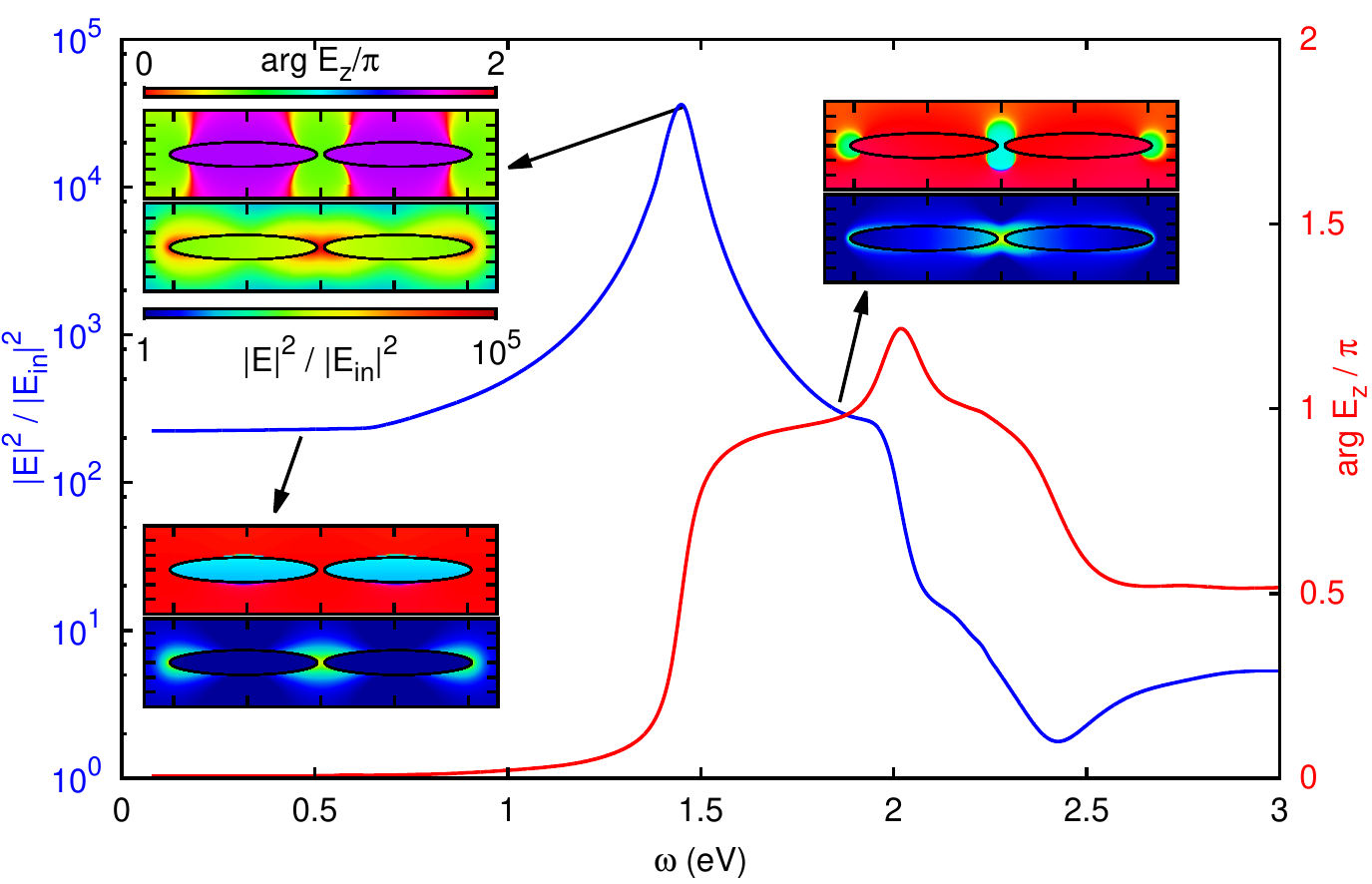}
\caption{Frequency-dependent intensity enhancement (blue) and phase shift of $E_z$ (red) at the center of the structure. The insets show the spatial dependence of enhancement and phase shift for three frequencies: In the quasistatic regime, close to the resonance peak, and on the blue side of the resonance.}
\label{fig:freqdep}
\end{figure}

The frequency-dependent response $\vec E_n(\vec r)$, shown in \autoref{fig:freqdep}, in principle contains all information about the system.
There is a dominant plasmon resonance at a photon energy of around $1.44\eV$ ($\lambda\approx 859\nm$), with a lifetime of $\approx\!7.5\fs$. The maximum intensity enhancement at the center between the ellipsoids is $\approx\!3\cdot10^4$. As the resonance is crossed, the phase jumps by $\pi$ as expected (illustrated by the dominant $z$-component of the field). When driving with a short (broad-band) pulse, the response will thus be in phase for the frequency components on the red side of the resonance, but out of phase on the blue side. This implies that a Fourier-limited few-cycle incoming pulse will produce a distorted and prolonged response in the center of the structure, essentially because the plasmon resonance is excited and keeps oscillating even after the driving pulse is over [see \autoref{fig:synth_fields}(a)]. It is this distortion that we wish to compensate in the following.
The insets in \autoref{fig:freqdep} show the spatial dependence of the plasmon response of the structure. The field is strongly enhanced at the sharp tips of the ellipsoids, with the largest enhancement in the narrow gap at the center. HHG will therefore be dominantly produced in this ``hot spot.''
Crucially, although not surprisingly, the phase of the response is spatially uniform within this hot spot for all relevant frequencies. Fourier synthesis of the incoming pulse will thus produce a uniform near-field temporal response, enabling control of strong-field processes.

\subsection{Isolated attosecond pulse generation}
To produce isolated attosecond pulses with amplitude gating, the driving laser pulse must only reach a sufficiently large electric field strength for a single cycle, and with the correct carrier-envelope-phase~\cite{Christov97,BraKra2000,ScrIvaKie2006,Corkum07,KraIva2009}. Once the frequency-dependent response is known, the incoming pulse can be chosen such that an arbitrary pulse shape is synthesized in the hot spot where HHG takes place. This could be achieved with standard pulse shapers that allow control over amplitude and phase of the separate frequency components of an incoming broadband pulse~\cite{BauBriSey1997,Wei2000}. The electric field in the structure is then given by
\begin{equation}\label{eq:filtered_fourier_synth}
\vec E(\vec r,t) = \sum_n c_n f_n \vec E_n(\vec r) e^{-i\omega_n t} + c.c.\,,
\end{equation}
where the complex amplitudes $c_n$ describe the unshaped pulse, while $f_n$ are complex numbers representing the pulse shaper ($0\leq |f_n| \leq1$). For concreteness, we choose $c_n$ to give an incoming broadband Gaussian pulse with a central energy of $1.448\eV$ ($\lambda\approx856\nm$) and a FWHM bandwidth of $0.55\eV$, corresponding to a FWHM duration of $3.33\fs$.
We simulate HHG by the synthesized near field using the strong-field approximation (SFA)~\cite{Lewenstein94,IvaBraBur1996}. As we focus on the controllability of the process, we do not currently include additional effects induced by the presence of the metal surface~\cite{HusImHer2011}.
The high harmonics are then spectrally filtered ($10\eV$ FWHM bandwidth centered at $100\eV$) to create an attosecond pulse, which could be characterized by, \eg attosecond streaking~\cite{ConTarSto1997,GouUibKie2004}, or alternatively all-optical methods which only require measurement of the emitted HHG radiation~\cite{RazSchAus2011,KimZhaShi2013}.

To optimize isolated attosecond pulse generation, we choose a fixed value of $2\%$ for the noise level, \ie the percentage of HHG power that is not in the main attosecond pulse, $N=1-P_{as}/P_{tot}$. Here, $P_{as}$ ($P_{tot}$) is the main attosecond pulse (total HHG) power.
Using the well-developed tools of optimal control,
the main attosecond pulse power $P_{as}$ is then maximized by numerically optimizing the coefficients $f_n$ in \autoref{eq:filtered_fourier_synth} \cite{*[{We use the Subplex algorithm from }] [{, as implemented in }] SUBPLEX, *NLOPT}. 

\begin{figure}[tb]
  \includegraphics[width=\linewidth]{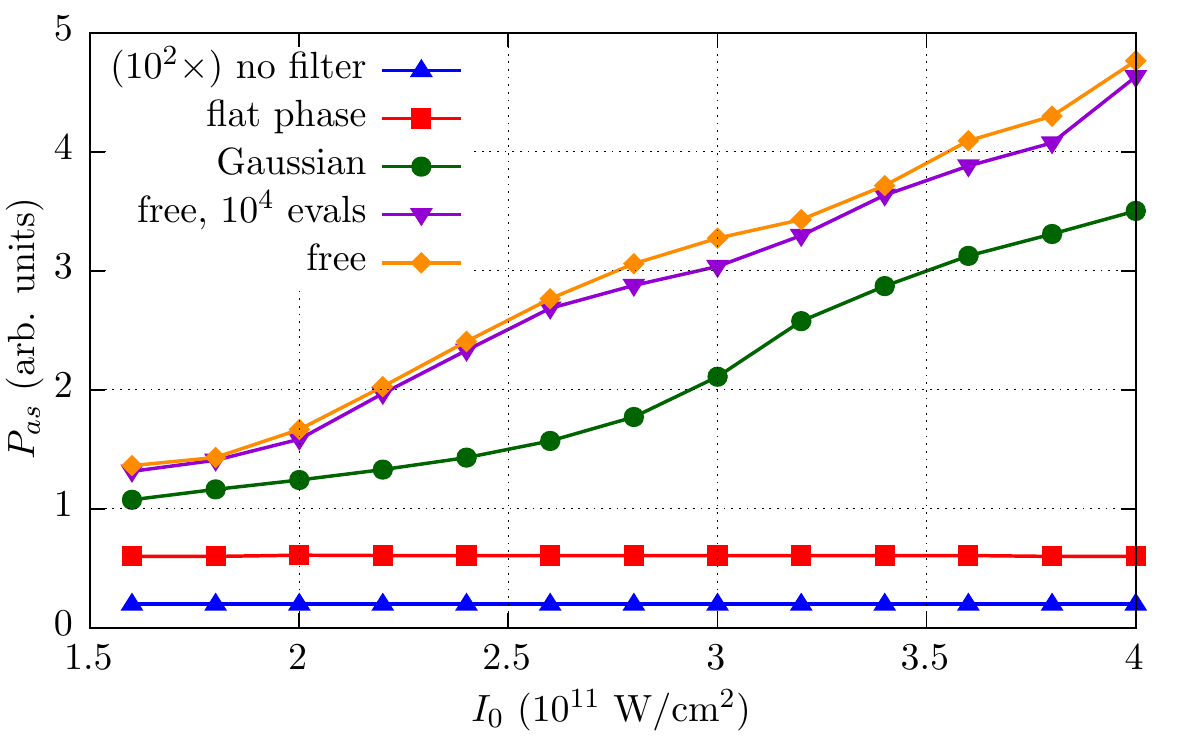}
  \caption{Optimized isolated attosecond pulse power $P_{as}$ for HHG from the plasmon-enhanced near-field transient for different filter settings.
  In all cases, the total intensity and CEP have been optimized. The noise $N$ is fixed to $2\%$, while the peak intensity $I_0$ of the incoming pulse is varied.
  For details, see text.}
  \label{fig:pulse_params}
\end{figure}

\begin{figure*}
  \includegraphics[width=0.328\linewidth]{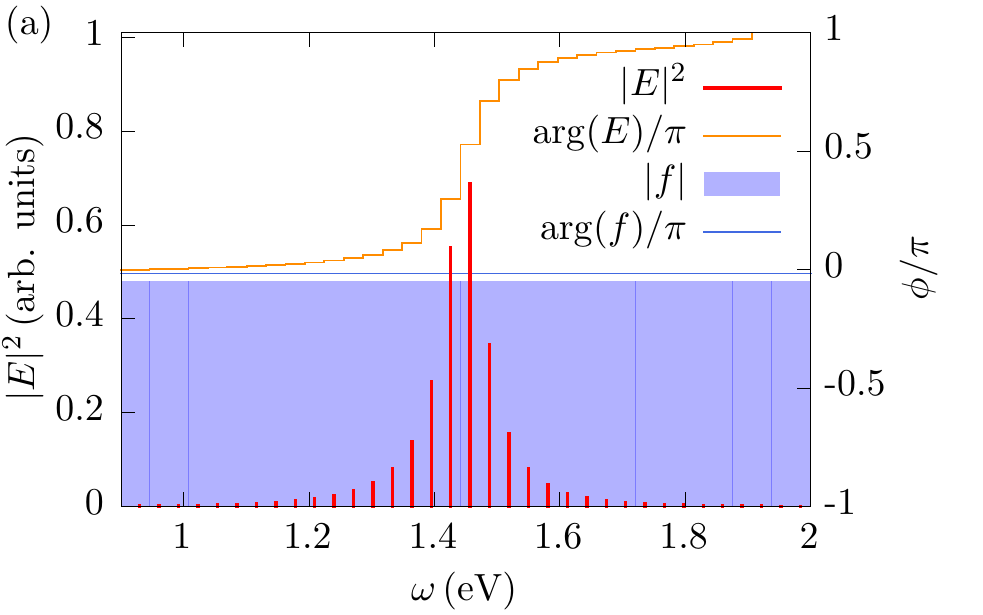}\hfill
  \includegraphics[width=0.328\linewidth]{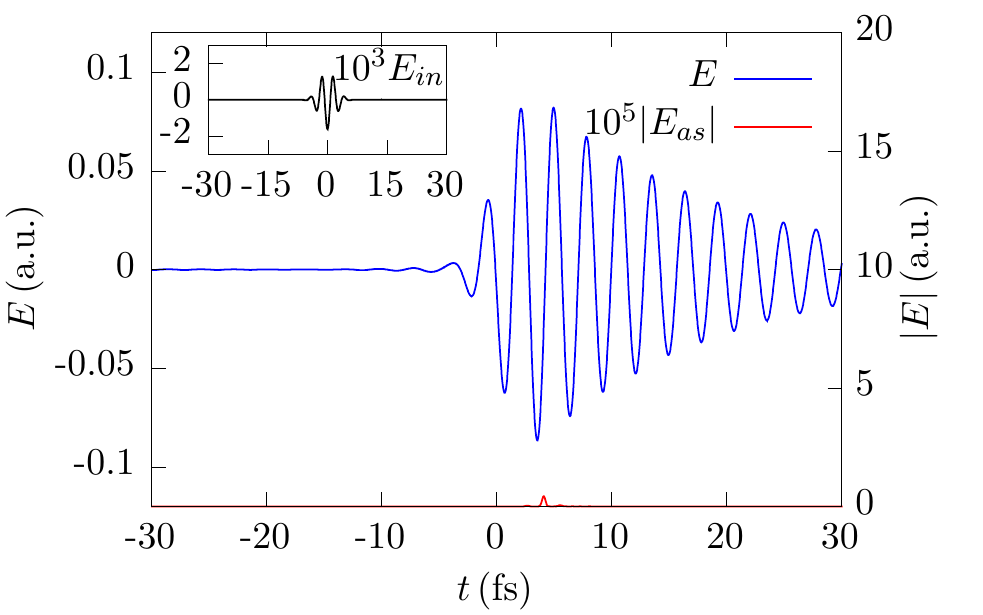}\hfill
  \includegraphics[width=0.328\linewidth]{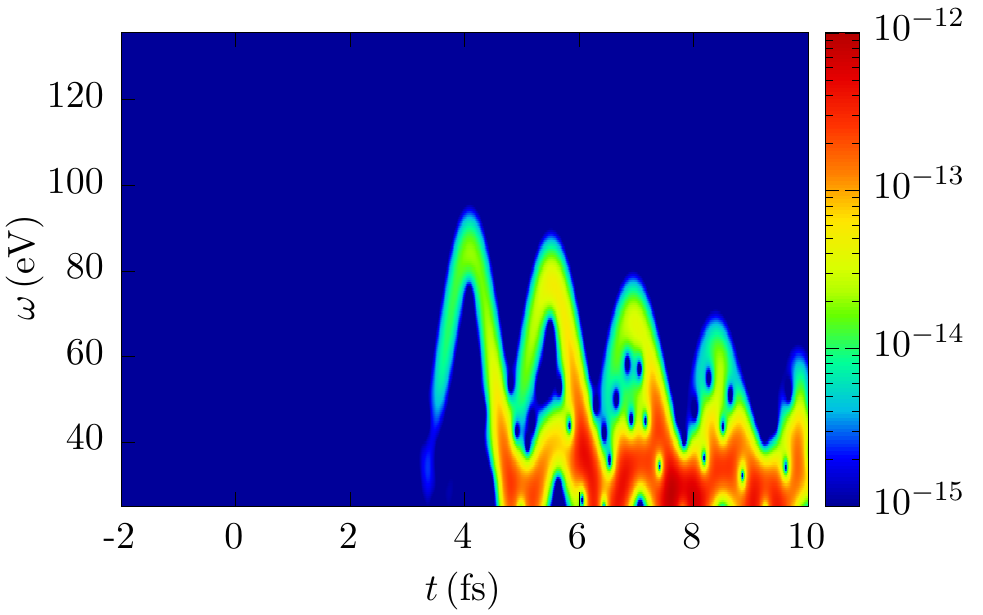}
  \includegraphics[width=0.328\linewidth]{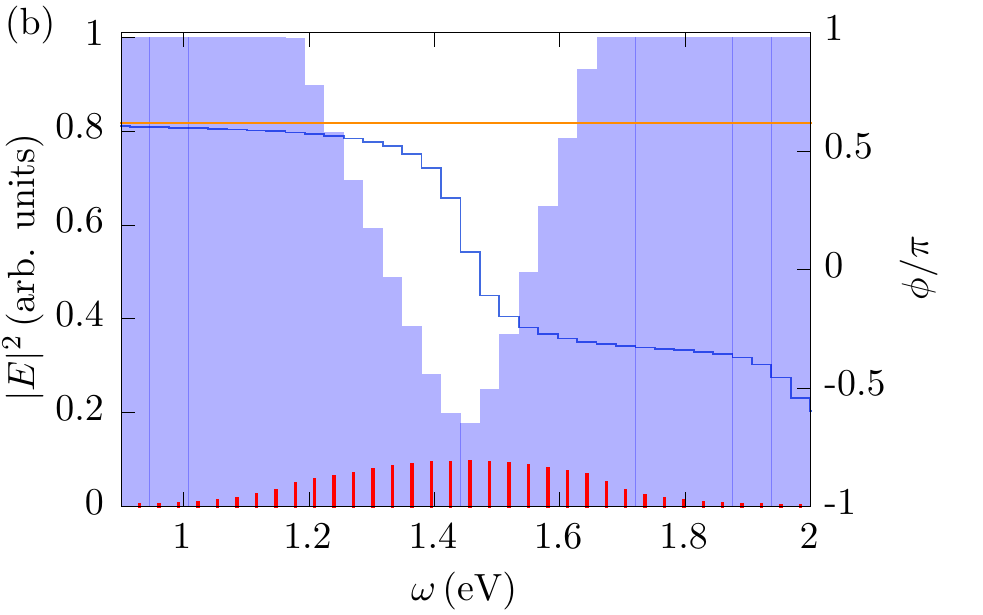}\hfill
  \includegraphics[width=0.328\linewidth]{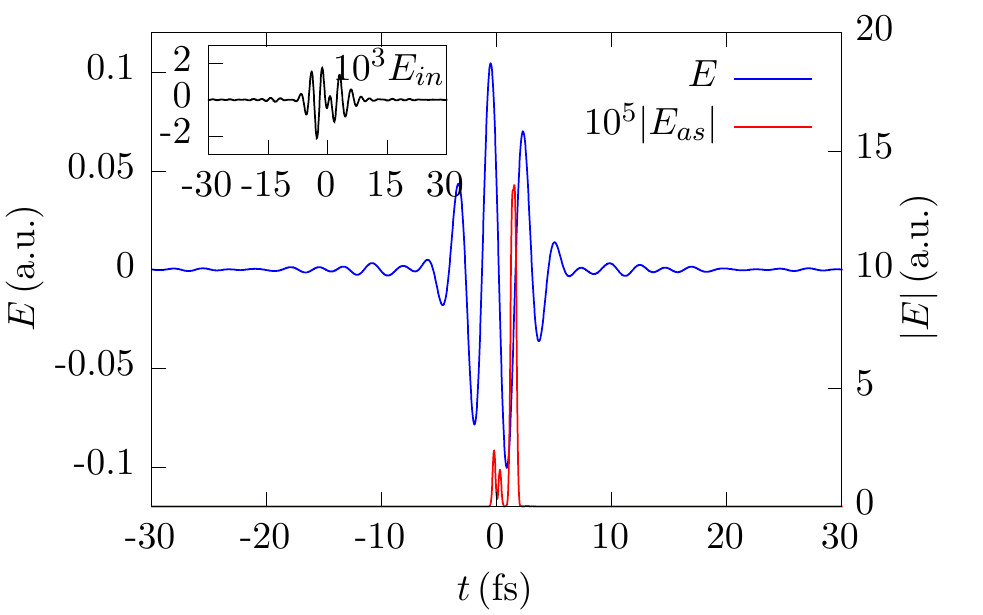}\hfill
  \includegraphics[width=0.328\linewidth]{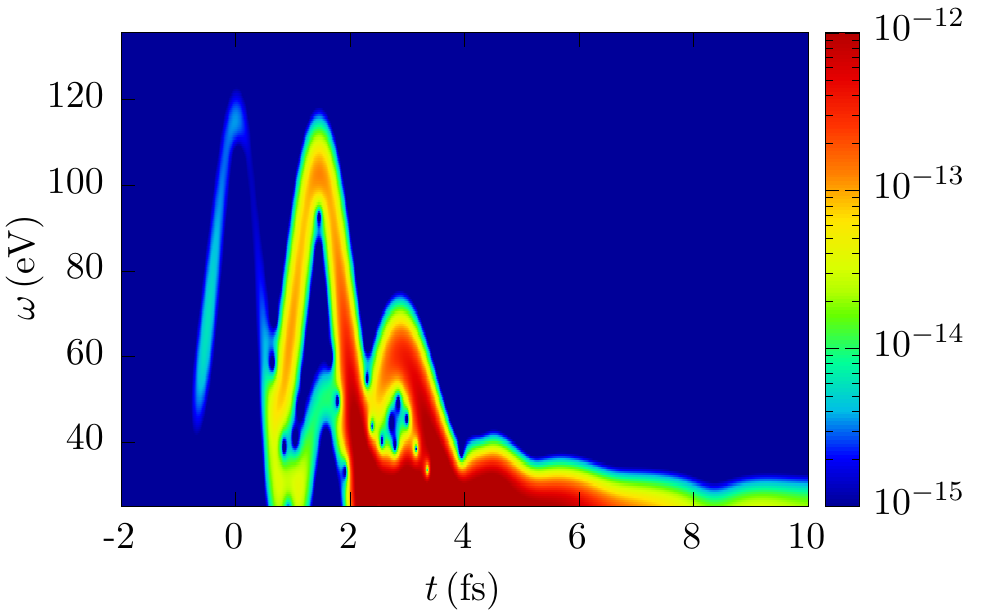}
  \includegraphics[width=0.328\linewidth]{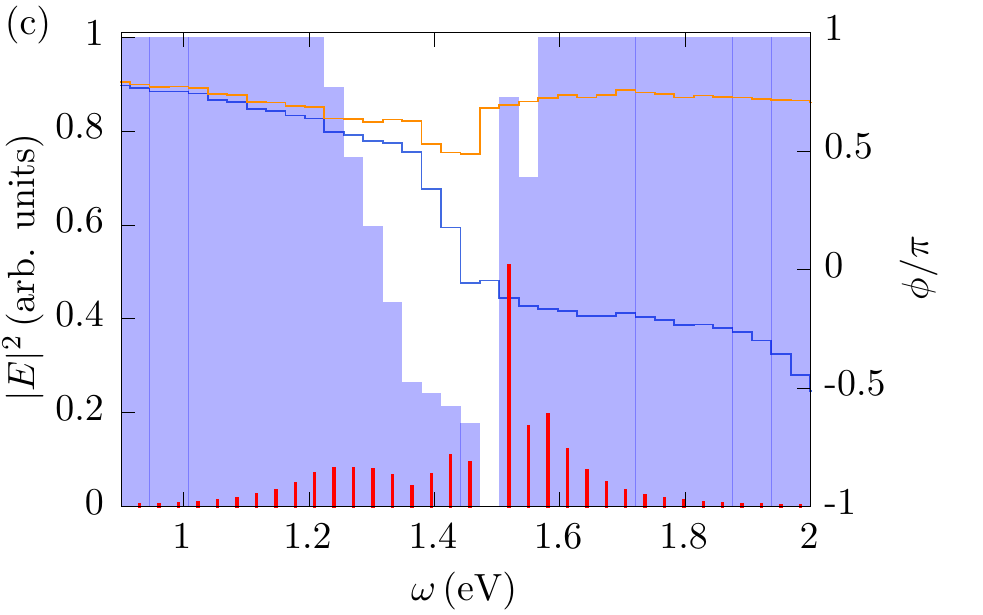}\hfill
  \includegraphics[width=0.328\linewidth]{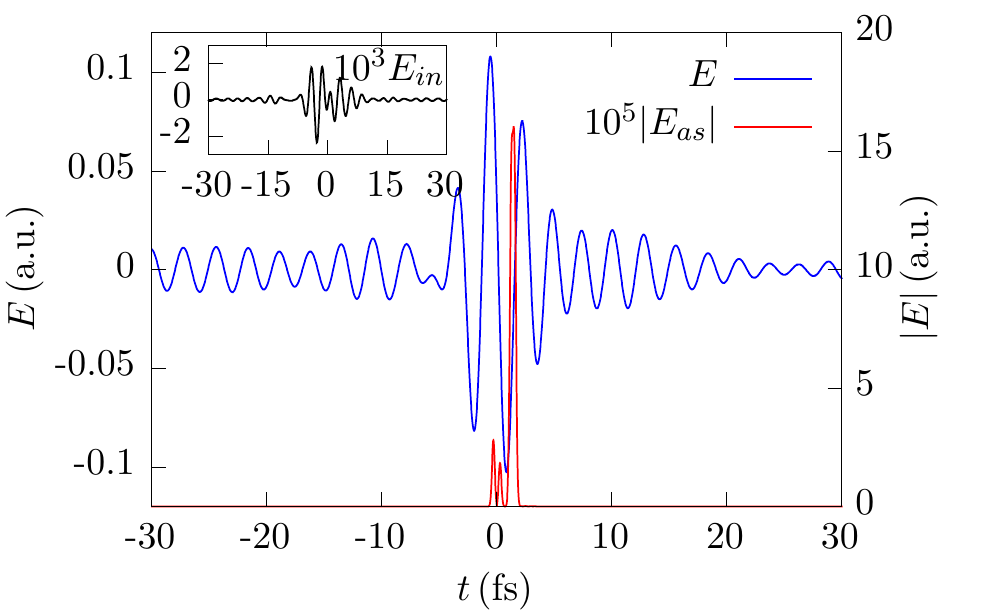}\hfill
  \includegraphics[width=0.328\linewidth]{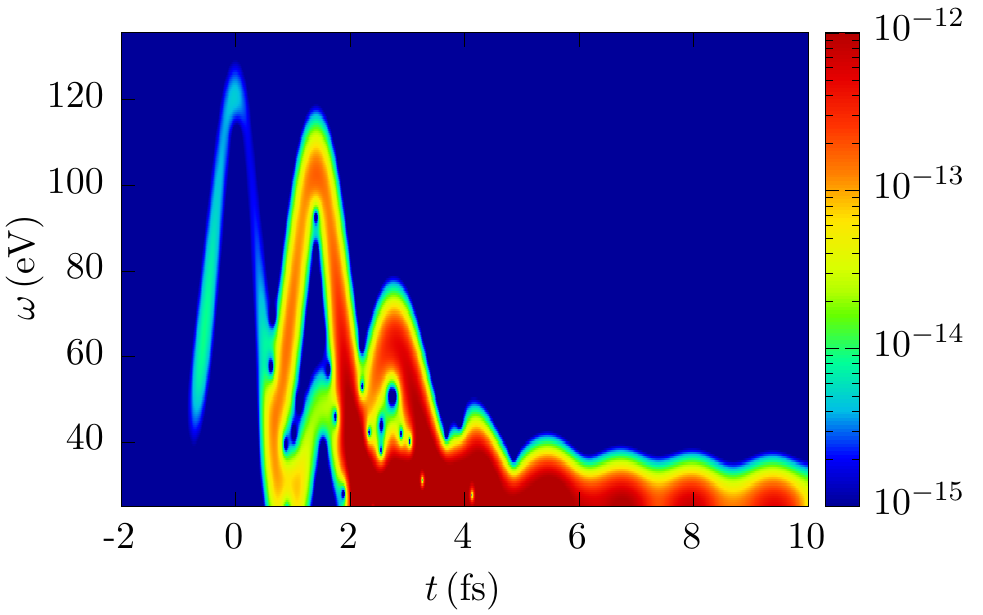}
\caption{Different synthesized pulses. In each row from left to right: Filter settings $f_n$ and electric field amplitudes $E_n$ (absolute value and phase are plotted separately), plasmon-enhanced electric field in the hot spot and filtered high harmonic radiation generated by this pulse (inset: incoming pulse), and time-frequency distribution of unfiltered high harmonic radiation.
The incoming pulse has a maximum intensity of $4\cdot10^{11}\Wcm$ before filtering.
The rows, from top to bottom, show:
(a) Fourier-limited incoming pulse with only amplitude and CEP adjusted to optimize single as pulse production in the hot spot.
(b) Manually optimized pulse chosen to produce a few-cycle Gaussian pulse in the hot spot as well as possible, amplitude and CEP optimized.
(c) Pulse obtained after numerical optimization (see text).
The numerically optimized pulse features a phase jump before the ``spike'' producing the attosecond pulse, effectively suppressing HHG from the previous half-cycle of the field.}
\label{fig:synth_fields}
\end{figure*}

In \autoref{fig:pulse_params}, we compare attosecond pulse generation from different synthesized near-field transients, \ie for different prescriptions of how to generate and optimize $f_n$. We optimize the total amplitude and CEP in all cases. If no further filtering of the incoming few-fs Gaussian pulse is performed, the aforementioned pulse distortion severely limits the achievable attosecond pulse power for acceptable noise levels. The associated temporal response and generated high harmonic radiation are shown in \autoref{fig:synth_fields}(a).

\vspace{-0.4cm}
\subsubsection{Manual optimization}
\vspace{-0.2cm}
 We first explore manual optimization of the near-field.
By adjusting just the \emph{phase} of the filter such that the near-field transient has a flat spectral phase (\ie is Fourier-transform limited),
attosecond pulse generation is significantly improved, and an increase in pulse power by more than two orders of magnitude is observed (red squares in \autoref{fig:pulse_params}).
For both the unfiltered and flat-phase near-field transients, the temporal shape and generated pulse power is independent of the incoming pulse intensity, as long as it is large enough to produce the necessary field strength for HHG. 

As a next step, not only the phases, but also the amplitudes of the spectral components can be adjusted to synthesize any desired near field.
The ``conventional'' choice for amplitude-gated isolated attosecond pulse generation is a few-cycle Gaussian pulse.
Since the plasmon resonance is typically more narrow-band than a few-cycle pulse (\ie the lifetime of the resonance is longer than the desired pulse), the incoming pulse must have a ``hole'' in the spectral distribution at the frequencies that are most strongly enhanced. The enhanced pulse then becomes nearly Gaussian, as shown in \autoref{fig:synth_fields}(b). For realistic incoming pulse intensities below the damage threshold of gold, the wings of the pulse will not follow a perfect Gaussian because of the limited field enhancement. More intense incoming pulses thus give more freedom to synthesize the desired transient, improving attosecond pulse generation (green circles in \autoref{fig:pulse_params}). For an incoming peak intensity of $4\cdot10^{11}\Wcm$, this manually optimized Gaussian pulse improves pulse power by another factor of five compared to just adjusting the phase.
At lower peak intensities, where the limited headroom in field enhancement prevents synthesis of a perfect Gaussian pulse,
it still provides an improvement of a factor of two or more.

\vspace{-0.4cm}
\subsubsection{Full optimization}
\vspace{-0.2cm}
The manual optimization as performed above has some drawbacks:
Because of the required spectral hole at the plasmon resonance, the achieved field amplification will be limited. In addition, it requires that the structure is perfectly characterized, which in reality will not always be the case. We thus additionally perform a completely free optimization of the filter parameters $f_n$, instead of prescribing any specific shape for the near-field transient.
This obviously needs more iterations to achieve convergence, requiring on the order of $10^5$ function evaluations. However, as the convergence is faster in the beginning and then slows down, the pulses obtained after $10^4$ function evaluations are almost as good (\cf\autoref{fig:pulse_params}). In an actual experiment, the time needed to optimize the pulse may be under a few tens of seconds, if an all-optical characterization of the pulses~\cite{RazSchAus2011,KimZhaShi2013} is applied (limited by the rate at which HHG spectra can be acquired and analyzed, which can reach the kHz regime).
For the same fixed noise level of $2\%$, this increases the pulse power significantly (by up to $75\%$) compared to the manually optimized Gaussian.

The exact shape of the fully optimized near-field transient depends quite sensitively on the parameters, while the optimized attosecond pulse power $P_{as}$ is relatively stable. In all cases, we found the same general features: The amplitude gating (just one half-cycle with enough intensity to produce the desired harmonic frequencies) is achieved by superposing a central (approximately Gaussian) ``spike'' with a longer background pulse, \cf\autoref{fig:synth_fields}(c). These two pulses switch from destructive to constructive interference within the width of the few-cycle pulse.
The field minimum from destructive interference effectively suppresses HHG in the half-cycle before the spike that generates the main attosecond pulse. This maximizes the single attosecond pulse generation efficiency while keeping noise low; consequently, this effect becomes even more pronounced if lower noise levels than shown here are chosen. In this way, the few-cycle pulse can be much broader in frequency than the plasmon resonance and ``boost'' its intensity for HHG by constructive interference with the longer narrow-band pulse. The long pulse by itself does not reach the intensity to produce high harmonic radiation at the $100\eV$ desired for the attosecond pulse, so its longer duration does not increase the noise level.

\section{Summary}\label{sec:summary}
To summarize, we have introduced a general approach to synthesize few-cycle nanoplasmonic near-fields and demonstrated its potential for the generation of isolated attosecond pulses. We have shown that the near-field transient can be synthesized to produce isolated attosecond pulses even in situations where the plasmon resonance is too long-lived to support generation of a strongly enhanced few-cycle near field. This can be achieved through straightforward manual optimization if the plasmon response is well-characterized. Furthermore, fully automated optimization leads to even better attosecond pulse generation by shutting off HHG for a half-cycle through destructive interference.

\begin{acknowledgments}
J.F.\ is grateful for support by the NSF through a grant to ITAMP and by the European Research Council under Grant No. 290981 (PLASMONANOQUANTA).
M.F.K.\ is grateful for support by the U.S.\ Department of Energy under DE-SC0008146 and DE-FG02-86ER13491, the BMBF via PhoNa, and the DFG via Kl-1439/4, and Kl-1439/5, and the Cluster of Excellence: Munich Center for Advanced Photonics (MAP).
M.T.H.R.\ is grateful for support by the Defense Advanced Research
Projects Agency (DARPA) under grant N66001-09-1-2070-DOD and by the 
AFOSR Multidisciplinary Research Program of the University Research 
Initiative (MURI) for Complex and Robust On-chip Nanophotonics under 
grant FA9550-09-1-0704.
\end{acknowledgments}


\begin{thebibliography}{49}%
\makeatletter
\providecommand \@ifxundefined [1]{%
 \@ifx{#1\undefined}
}%
\providecommand \@ifnum [1]{%
 \ifnum #1\expandafter \@firstoftwo
 \else \expandafter \@secondoftwo
 \fi
}%
\providecommand \@ifx [1]{%
 \ifx #1\expandafter \@firstoftwo
 \else \expandafter \@secondoftwo
 \fi
}%
\providecommand \natexlab [1]{#1}%
\providecommand \enquote  [1]{``#1''}%
\providecommand \bibnamefont  [1]{#1}%
\providecommand \bibfnamefont [1]{#1}%
\providecommand \citenamefont [1]{#1}%
\providecommand \href@noop [0]{\@secondoftwo}%
\providecommand \href [0]{\begingroup \@sanitize@url \@href}%
\providecommand \@href[1]{\@@startlink{#1}\@@href}%
\providecommand \@@href[1]{\endgroup#1\@@endlink}%
\providecommand \@sanitize@url [0]{\catcode `\\12\catcode `\$12\catcode
  `\&12\catcode `\#12\catcode `\^12\catcode `\_12\catcode `\%12\relax}%
\providecommand \@@startlink[1]{}%
\providecommand \@@endlink[0]{}%
\providecommand \url  [0]{\begingroup\@sanitize@url \@url }%
\providecommand \@url [1]{\endgroup\@href {#1}{\urlprefix }}%
\providecommand \urlprefix  [0]{URL }%
\providecommand \Eprint [0]{\href }%
\providecommand \doibase [0]{http://dx.doi.org/}%
\providecommand \selectlanguage [0]{\@gobble}%
\providecommand \bibinfo  [0]{\@secondoftwo}%
\providecommand \bibfield  [0]{\@secondoftwo}%
\providecommand \translation [1]{[#1]}%
\providecommand \BibitemOpen [0]{}%
\providecommand \bibitemStop [0]{}%
\providecommand \bibitemNoStop [0]{.\EOS\space}%
\providecommand \EOS [0]{\spacefactor3000\relax}%
\providecommand \BibitemShut  [1]{\csname bibitem#1\endcsname}%
\let\auto@bib@innerbib\@empty
%</preamble>
\bibitem [{\citenamefont {Krausz}\ and\ \citenamefont
  {Ivanov}(2009)}]{KraIva2009}%
  \BibitemOpen
  \bibfield  {author} {\bibinfo {author} {\bibfnamefont {F.}~\bibnamefont
  {Krausz}}\ and\ \bibinfo {author} {\bibfnamefont {M.}~\bibnamefont
  {Ivanov}},\ }\href {\doibase 10.1103/revmodphys.81.163} {\bibfield  {journal}
  {\bibinfo  {journal} {Rev. Mod. Phys}\ }\textbf {\bibinfo {volume} {81}},\
  \bibinfo {pages} {163} (\bibinfo {year} {2009})}\BibitemShut {NoStop}%
\bibitem [{\citenamefont {Goulielmakis}\ \emph {et~al.}(2008)\citenamefont
  {Goulielmakis}, \citenamefont {Schultze}, \citenamefont {Hofstetter},
  \citenamefont {Yakovlev}, \citenamefont {Gagnon}, \citenamefont {Uiberacker},
  \citenamefont {Aquila}, \citenamefont {Gullikson}, \citenamefont {Attwood},
  \citenamefont {Kienberger}, \citenamefont {Krausz},\ and\ \citenamefont
  {Kleineberg}}]{GouSchHof2008}%
  \BibitemOpen
  \bibfield  {author} {\bibinfo {author} {\bibfnamefont {E.}~\bibnamefont
  {Goulielmakis}}, \bibinfo {author} {\bibfnamefont {M.}~\bibnamefont
  {Schultze}}, \bibinfo {author} {\bibfnamefont {M.}~\bibnamefont
  {Hofstetter}}, \bibinfo {author} {\bibfnamefont {V.~S.}\ \bibnamefont
  {Yakovlev}}, \bibinfo {author} {\bibfnamefont {J.}~\bibnamefont {Gagnon}},
  \bibinfo {author} {\bibfnamefont {M.}~\bibnamefont {Uiberacker}}, \bibinfo
  {author} {\bibfnamefont {A.~L.}\ \bibnamefont {Aquila}}, \bibinfo {author}
  {\bibfnamefont {E.~M.}\ \bibnamefont {Gullikson}}, \bibinfo {author}
  {\bibfnamefont {D.~T.}\ \bibnamefont {Attwood}}, \bibinfo {author}
  {\bibfnamefont {R.}~\bibnamefont {Kienberger}}, \bibinfo {author}
  {\bibfnamefont {F.}~\bibnamefont {Krausz}}, \ and\ \bibinfo {author}
  {\bibfnamefont {U.}~\bibnamefont {Kleineberg}},\ }\href {\doibase
  10.1126/science.1157846} {\bibfield  {journal} {\bibinfo  {journal}
  {Science}\ }\textbf {\bibinfo {volume} {320}},\ \bibinfo {pages} {1614}
  (\bibinfo {year} {2008})}\BibitemShut {NoStop}%
\bibitem [{\citenamefont {Zhao}\ \emph {et~al.}(2012)\citenamefont {Zhao},
  \citenamefont {Zhang}, \citenamefont {Chini}, \citenamefont {Wu},
  \citenamefont {Wang},\ and\ \citenamefont {Chang}}]{ZhaZhaChi2012}%
  \BibitemOpen
  \bibfield  {author} {\bibinfo {author} {\bibfnamefont {K.}~\bibnamefont
  {Zhao}}, \bibinfo {author} {\bibfnamefont {Q.}~\bibnamefont {Zhang}},
  \bibinfo {author} {\bibfnamefont {M.}~\bibnamefont {Chini}}, \bibinfo
  {author} {\bibfnamefont {Y.}~\bibnamefont {Wu}}, \bibinfo {author}
  {\bibfnamefont {X.}~\bibnamefont {Wang}}, \ and\ \bibinfo {author}
  {\bibfnamefont {Z.}~\bibnamefont {Chang}},\ }\href {\doibase
  10.1364/ol.37.003891} {\bibfield  {journal} {\bibinfo  {journal} {Opt.
  Lett.}\ }\textbf {\bibinfo {volume} {37}},\ \bibinfo {pages} {3891} (\bibinfo
  {year} {2012})}\BibitemShut {NoStop}%
\bibitem [{\citenamefont {Schultze}\ \emph {et~al.}(2010)\citenamefont
  {Schultze}, \citenamefont {Fiess}, \citenamefont {Karpowicz}, \citenamefont
  {Gagnon}, \citenamefont {Korbman}, \citenamefont {Hofstetter}, \citenamefont
  {Neppl}, \citenamefont {Cavalieri}, \citenamefont {Komninos}, \citenamefont
  {Mercouris}, \citenamefont {Nicolaides}, \citenamefont {Pazourek},
  \citenamefont {Nagele}, \citenamefont {Feist}, \citenamefont
  {Burgd\"{o}rfer}, \citenamefont {Azzeer}, \citenamefont {Ernstorfer},
  \citenamefont {Kienberger}, \citenamefont {Kleineberg}, \citenamefont
  {Goulielmakis}, \citenamefont {Krausz},\ and\ \citenamefont
  {Yakovlev}}]{SchFieKar2010}%
  \BibitemOpen
  \bibfield  {author} {\bibinfo {author} {\bibfnamefont {M.}~\bibnamefont
  {Schultze}}, \bibinfo {author} {\bibfnamefont {M.}~\bibnamefont {Fiess}},
  \bibinfo {author} {\bibfnamefont {N.}~\bibnamefont {Karpowicz}}, \bibinfo
  {author} {\bibfnamefont {J.}~\bibnamefont {Gagnon}}, \bibinfo {author}
  {\bibfnamefont {M.}~\bibnamefont {Korbman}}, \bibinfo {author} {\bibfnamefont
  {M.}~\bibnamefont {Hofstetter}}, \bibinfo {author} {\bibfnamefont
  {S.}~\bibnamefont {Neppl}}, \bibinfo {author} {\bibfnamefont {A.~L.}\
  \bibnamefont {Cavalieri}}, \bibinfo {author} {\bibfnamefont {Y.}~\bibnamefont
  {Komninos}}, \bibinfo {author} {\bibfnamefont {T.}~\bibnamefont {Mercouris}},
  \bibinfo {author} {\bibfnamefont {C.~A.}\ \bibnamefont {Nicolaides}},
  \bibinfo {author} {\bibfnamefont {R.}~\bibnamefont {Pazourek}}, \bibinfo
  {author} {\bibfnamefont {S.}~\bibnamefont {Nagele}}, \bibinfo {author}
  {\bibfnamefont {J.}~\bibnamefont {Feist}}, \bibinfo {author} {\bibfnamefont
  {J.}~\bibnamefont {Burgd\"{o}rfer}}, \bibinfo {author} {\bibfnamefont
  {A.~M.}\ \bibnamefont {Azzeer}}, \bibinfo {author} {\bibfnamefont
  {R.}~\bibnamefont {Ernstorfer}}, \bibinfo {author} {\bibfnamefont
  {R.}~\bibnamefont {Kienberger}}, \bibinfo {author} {\bibfnamefont
  {U.}~\bibnamefont {Kleineberg}}, \bibinfo {author} {\bibfnamefont
  {E.}~\bibnamefont {Goulielmakis}}, \bibinfo {author} {\bibfnamefont
  {F.}~\bibnamefont {Krausz}}, \ and\ \bibinfo {author} {\bibfnamefont {V.~S.}\
  \bibnamefont {Yakovlev}},\ }\href {\doibase 10.1126/science.1189401}
  {\bibfield  {journal} {\bibinfo  {journal} {Science}\ }\textbf {\bibinfo
  {volume} {328}},\ \bibinfo {pages} {1658} (\bibinfo {year}
  {2010})}\BibitemShut {NoStop}%
\bibitem [{\citenamefont {Goulielmakis}\ \emph {et~al.}(2010)\citenamefont
  {Goulielmakis}, \citenamefont {Loh}, \citenamefont {Wirth}, \citenamefont
  {Santra}, \citenamefont {Rohringer}, \citenamefont {Yakovlev}, \citenamefont
  {Zherebtsov}, \citenamefont {Pfeifer}, \citenamefont {Azzeer}, \citenamefont
  {Kling}, \citenamefont {Leone},\ and\ \citenamefont
  {Krausz}}]{GouLohWir2010}%
  \BibitemOpen
  \bibfield  {author} {\bibinfo {author} {\bibfnamefont {E.}~\bibnamefont
  {Goulielmakis}}, \bibinfo {author} {\bibfnamefont {Z.-H.}\ \bibnamefont
  {Loh}}, \bibinfo {author} {\bibfnamefont {A.}~\bibnamefont {Wirth}}, \bibinfo
  {author} {\bibfnamefont {R.}~\bibnamefont {Santra}}, \bibinfo {author}
  {\bibfnamefont {N.}~\bibnamefont {Rohringer}}, \bibinfo {author}
  {\bibfnamefont {V.~S.}\ \bibnamefont {Yakovlev}}, \bibinfo {author}
  {\bibfnamefont {S.}~\bibnamefont {Zherebtsov}}, \bibinfo {author}
  {\bibfnamefont {T.}~\bibnamefont {Pfeifer}}, \bibinfo {author} {\bibfnamefont
  {A.~M.}\ \bibnamefont {Azzeer}}, \bibinfo {author} {\bibfnamefont {M.~F.}\
  \bibnamefont {Kling}}, \bibinfo {author} {\bibfnamefont {S.~R.}\ \bibnamefont
  {Leone}}, \ and\ \bibinfo {author} {\bibfnamefont {F.}~\bibnamefont
  {Krausz}},\ }\href {\doibase 10.1038/nature09212} {\bibfield  {journal}
  {\bibinfo  {journal} {Nature}\ }\textbf {\bibinfo {volume} {466}},\ \bibinfo
  {pages} {739} (\bibinfo {year} {2010})}\BibitemShut {NoStop}%
\bibitem [{\citenamefont {Kl\"{u}nder}\ \emph {et~al.}(2011)\citenamefont
  {Kl\"{u}nder}, \citenamefont {Dahlstr\"{o}m}, \citenamefont {Gisselbrecht},
  \citenamefont {Fordell}, \citenamefont {Swoboda}, \citenamefont {Gu\'{e}not},
  \citenamefont {Johnsson}, \citenamefont {Caillat}, \citenamefont
  {Mauritsson}, \citenamefont {Maquet}, \citenamefont {Ta\"{i}eb},\ and\
  \citenamefont {L'Huillier}}]{KluDahGis2011}%
  \BibitemOpen
  \bibfield  {author} {\bibinfo {author} {\bibfnamefont {K.}~\bibnamefont
  {Kl\"{u}nder}}, \bibinfo {author} {\bibfnamefont {J.~M.}\ \bibnamefont
  {Dahlstr\"{o}m}}, \bibinfo {author} {\bibfnamefont {M.}~\bibnamefont
  {Gisselbrecht}}, \bibinfo {author} {\bibfnamefont {T.}~\bibnamefont
  {Fordell}}, \bibinfo {author} {\bibfnamefont {M.}~\bibnamefont {Swoboda}},
  \bibinfo {author} {\bibfnamefont {D.}~\bibnamefont {Gu\'{e}not}}, \bibinfo
  {author} {\bibfnamefont {P.}~\bibnamefont {Johnsson}}, \bibinfo {author}
  {\bibfnamefont {J.}~\bibnamefont {Caillat}}, \bibinfo {author} {\bibfnamefont
  {J.}~\bibnamefont {Mauritsson}}, \bibinfo {author} {\bibfnamefont
  {A.}~\bibnamefont {Maquet}}, \bibinfo {author} {\bibfnamefont
  {R.}~\bibnamefont {Ta\"{i}eb}}, \ and\ \bibinfo {author} {\bibfnamefont
  {A.}~\bibnamefont {L'Huillier}},\ }\href {\doibase
  10.1103/physrevlett.106.143002} {\bibfield  {journal} {\bibinfo  {journal}
  {Phys. Rev. Lett.}\ }\textbf {\bibinfo {volume} {106}},\ \bibinfo {pages}
  {143002} (\bibinfo {year} {2011})}\BibitemShut {NoStop}%
\bibitem [{\citenamefont {Kling}\ \emph {et~al.}(2006)\citenamefont {Kling},
  \citenamefont {Siedschlag}, \citenamefont {Verhoef}, \citenamefont {Khan},
  \citenamefont {Schultze}, \citenamefont {Uphues}, \citenamefont {Ni},
  \citenamefont {Uiberacker}, \citenamefont {Drescher}, \citenamefont
  {Krausz},\ and\ \citenamefont {Vrakking}}]{KliSieVer2006}%
  \BibitemOpen
  \bibfield  {author} {\bibinfo {author} {\bibfnamefont {M.~F.}\ \bibnamefont
  {Kling}}, \bibinfo {author} {\bibfnamefont {C.}~\bibnamefont {Siedschlag}},
  \bibinfo {author} {\bibfnamefont {A.~J.}\ \bibnamefont {Verhoef}}, \bibinfo
  {author} {\bibfnamefont {J.~I.}\ \bibnamefont {Khan}}, \bibinfo {author}
  {\bibfnamefont {M.}~\bibnamefont {Schultze}}, \bibinfo {author}
  {\bibfnamefont {T.}~\bibnamefont {Uphues}}, \bibinfo {author} {\bibfnamefont
  {Y.}~\bibnamefont {Ni}}, \bibinfo {author} {\bibfnamefont {M.}~\bibnamefont
  {Uiberacker}}, \bibinfo {author} {\bibfnamefont {M.}~\bibnamefont
  {Drescher}}, \bibinfo {author} {\bibfnamefont {F.}~\bibnamefont {Krausz}}, \
  and\ \bibinfo {author} {\bibfnamefont {M.~J.~J.}\ \bibnamefont {Vrakking}},\
  }\href {\doibase 10.1126/science.1126259} {\bibfield  {journal} {\bibinfo
  {journal} {Science}\ }\textbf {\bibinfo {volume} {312}},\ \bibinfo {pages}
  {246} (\bibinfo {year} {2006})}\BibitemShut {NoStop}%
\bibitem [{\citenamefont {Sansone}\ \emph {et~al.}(2010)\citenamefont
  {Sansone}, \citenamefont {Kelkensberg}, \citenamefont {Perez-Torres},
  \citenamefont {Morales}, \citenamefont {Kling}, \citenamefont {Siu},
  \citenamefont {Ghafur}, \citenamefont {Johnsson}, \citenamefont {Swoboda},
  \citenamefont {Benedetti}, \citenamefont {Ferrari}, \citenamefont {Lepine},
  \citenamefont {Sanz-Vicario}, \citenamefont {Zherebtsov}, \citenamefont
  {Znakovskaya}, \citenamefont {L'Huillier}, \citenamefont {Ivanov},
  \citenamefont {Nisoli}, \citenamefont {Mart\'{\i}n},\ and\ \citenamefont
  {Vrakking}}]{SanKelPer2010}%
  \BibitemOpen
  \bibfield  {author} {\bibinfo {author} {\bibfnamefont {G.}~\bibnamefont
  {Sansone}}, \bibinfo {author} {\bibfnamefont {F.}~\bibnamefont
  {Kelkensberg}}, \bibinfo {author} {\bibfnamefont {J.~F.}\ \bibnamefont
  {Perez-Torres}}, \bibinfo {author} {\bibfnamefont {F.}~\bibnamefont
  {Morales}}, \bibinfo {author} {\bibfnamefont {M.~F.}\ \bibnamefont {Kling}},
  \bibinfo {author} {\bibfnamefont {W.}~\bibnamefont {Siu}}, \bibinfo {author}
  {\bibfnamefont {O.}~\bibnamefont {Ghafur}}, \bibinfo {author} {\bibfnamefont
  {P.}~\bibnamefont {Johnsson}}, \bibinfo {author} {\bibfnamefont
  {M.}~\bibnamefont {Swoboda}}, \bibinfo {author} {\bibfnamefont
  {E.}~\bibnamefont {Benedetti}}, \bibinfo {author} {\bibfnamefont
  {F.}~\bibnamefont {Ferrari}}, \bibinfo {author} {\bibfnamefont
  {F.}~\bibnamefont {Lepine}}, \bibinfo {author} {\bibfnamefont {J.~L.}\
  \bibnamefont {Sanz-Vicario}}, \bibinfo {author} {\bibfnamefont
  {S.}~\bibnamefont {Zherebtsov}}, \bibinfo {author} {\bibfnamefont
  {I.}~\bibnamefont {Znakovskaya}}, \bibinfo {author} {\bibfnamefont
  {A.}~\bibnamefont {L'Huillier}}, \bibinfo {author} {\bibnamefont {Ivanov}},
  \bibinfo {author} {\bibfnamefont {M.}~\bibnamefont {Nisoli}}, \bibinfo
  {author} {\bibfnamefont {F.}~\bibnamefont {Mart\'{\i}n}}, \ and\ \bibinfo
  {author} {\bibfnamefont {M.~J.~J.}\ \bibnamefont {Vrakking}},\ }\href
  {\doibase 10.1038/nature09084} {\bibfield  {journal} {\bibinfo  {journal}
  {Nature}\ }\textbf {\bibinfo {volume} {465}},\ \bibinfo {pages} {763}
  (\bibinfo {year} {2010})}\BibitemShut {NoStop}%
\bibitem [{\citenamefont {Apolonski}\ \emph {et~al.}(2004)\citenamefont
  {Apolonski}, \citenamefont {Dombi}, \citenamefont {Paulus}, \citenamefont
  {Kakehata}, \citenamefont {Holzwarth}, \citenamefont {Th}, \citenamefont
  {Ch}, \citenamefont {Torizuka}, \citenamefont {Burgd\"{o}rfer}, \citenamefont
  {H\"{a}nsch},\ and\ \citenamefont {Krausz}}]{ApoDomPau2004}%
  \BibitemOpen
  \bibfield  {author} {\bibinfo {author} {\bibfnamefont {A.}~\bibnamefont
  {Apolonski}}, \bibinfo {author} {\bibfnamefont {P.}~\bibnamefont {Dombi}},
  \bibinfo {author} {\bibfnamefont {G.~G.}\ \bibnamefont {Paulus}}, \bibinfo
  {author} {\bibfnamefont {M.}~\bibnamefont {Kakehata}}, \bibinfo {author}
  {\bibfnamefont {R.}~\bibnamefont {Holzwarth}}, \bibinfo {author}
  {\bibnamefont {Th}}, \bibinfo {author} {\bibnamefont {Ch}}, \bibinfo {author}
  {\bibfnamefont {K.}~\bibnamefont {Torizuka}}, \bibinfo {author}
  {\bibfnamefont {J.}~\bibnamefont {Burgd\"{o}rfer}}, \bibinfo {author}
  {\bibfnamefont {T.~W.}\ \bibnamefont {H\"{a}nsch}}, \ and\ \bibinfo {author}
  {\bibfnamefont {F.}~\bibnamefont {Krausz}},\ }\href {\doibase
  10.1103/physrevlett.92.073902} {\bibfield  {journal} {\bibinfo  {journal}
  {Phys. Rev. Lett.}\ }\textbf {\bibinfo {volume} {92}},\ \bibinfo {pages}
  {073902} (\bibinfo {year} {2004})}\BibitemShut {NoStop}%
\bibitem [{\citenamefont {Cavalieri}\ \emph {et~al.}(2007)\citenamefont
  {Cavalieri}, \citenamefont {M\"{u}ller}, \citenamefont {Uphues},
  \citenamefont {Yakovlev}, \citenamefont {Baltu\v{s}ka}, \citenamefont
  {Horvath}, \citenamefont {Schmidt}, \citenamefont {Bl\"{u}mel}, \citenamefont
  {Holzwarth}, \citenamefont {Hendel}, \citenamefont {Drescher}, \citenamefont
  {Kleineberg}, \citenamefont {Echenique}, \citenamefont {Kienberger},
  \citenamefont {Krausz},\ and\ \citenamefont {Heinzmann}}]{CavMueUph2007}%
  \BibitemOpen
  \bibfield  {author} {\bibinfo {author} {\bibfnamefont {A.~L.}\ \bibnamefont
  {Cavalieri}}, \bibinfo {author} {\bibfnamefont {N.}~\bibnamefont
  {M\"{u}ller}}, \bibinfo {author} {\bibfnamefont {T.}~\bibnamefont {Uphues}},
  \bibinfo {author} {\bibfnamefont {V.~S.}\ \bibnamefont {Yakovlev}}, \bibinfo
  {author} {\bibfnamefont {A.}~\bibnamefont {Baltu\v{s}ka}}, \bibinfo {author}
  {\bibfnamefont {B.}~\bibnamefont {Horvath}}, \bibinfo {author} {\bibfnamefont
  {B.}~\bibnamefont {Schmidt}}, \bibinfo {author} {\bibfnamefont
  {L.}~\bibnamefont {Bl\"{u}mel}}, \bibinfo {author} {\bibfnamefont
  {R.}~\bibnamefont {Holzwarth}}, \bibinfo {author} {\bibfnamefont
  {S.}~\bibnamefont {Hendel}}, \bibinfo {author} {\bibfnamefont
  {M.}~\bibnamefont {Drescher}}, \bibinfo {author} {\bibfnamefont
  {U.}~\bibnamefont {Kleineberg}}, \bibinfo {author} {\bibfnamefont {P.~M.}\
  \bibnamefont {Echenique}}, \bibinfo {author} {\bibfnamefont {R.}~\bibnamefont
  {Kienberger}}, \bibinfo {author} {\bibfnamefont {F.}~\bibnamefont {Krausz}},
  \ and\ \bibinfo {author} {\bibfnamefont {U.}~\bibnamefont {Heinzmann}},\
  }\href {\doibase 10.1038/nature06229} {\bibfield  {journal} {\bibinfo
  {journal} {Nature}\ }\textbf {\bibinfo {volume} {449}},\ \bibinfo {pages}
  {1029} (\bibinfo {year} {2007})}\BibitemShut {NoStop}%
\bibitem [{\citenamefont {Zherebtsov}\ \emph {et~al.}(2011)\citenamefont
  {Zherebtsov}, \citenamefont {Fennel}, \citenamefont {Plenge}, \citenamefont
  {Antonsson}, \citenamefont {Znakovskaya}, \citenamefont {Wirth},
  \citenamefont {Herrwerth}, \citenamefont {Suszmann}, \citenamefont {Peltz},
  \citenamefont {Ahmad}, \citenamefont {Trushin}, \citenamefont {Pervak},
  \citenamefont {Karsch}, \citenamefont {Vrakking}, \citenamefont {Langer},
  \citenamefont {Graf}, \citenamefont {Stockman}, \citenamefont {Krausz},
  \citenamefont {Ruhl},\ and\ \citenamefont {Kling}}]{ZheFenPle2011}%
  \BibitemOpen
  \bibfield  {author} {\bibinfo {author} {\bibfnamefont {S.}~\bibnamefont
  {Zherebtsov}}, \bibinfo {author} {\bibfnamefont {T.}~\bibnamefont {Fennel}},
  \bibinfo {author} {\bibfnamefont {J.}~\bibnamefont {Plenge}}, \bibinfo
  {author} {\bibfnamefont {E.}~\bibnamefont {Antonsson}}, \bibinfo {author}
  {\bibfnamefont {I.}~\bibnamefont {Znakovskaya}}, \bibinfo {author}
  {\bibfnamefont {A.}~\bibnamefont {Wirth}}, \bibinfo {author} {\bibfnamefont
  {O.}~\bibnamefont {Herrwerth}}, \bibinfo {author} {\bibfnamefont
  {F.}~\bibnamefont {Suszmann}}, \bibinfo {author} {\bibfnamefont
  {C.}~\bibnamefont {Peltz}}, \bibinfo {author} {\bibfnamefont
  {I.}~\bibnamefont {Ahmad}}, \bibinfo {author} {\bibfnamefont {S.~A.}\
  \bibnamefont {Trushin}}, \bibinfo {author} {\bibfnamefont {V.}~\bibnamefont
  {Pervak}}, \bibinfo {author} {\bibfnamefont {S.}~\bibnamefont {Karsch}},
  \bibinfo {author} {\bibfnamefont {M.~J.~J.}\ \bibnamefont {Vrakking}},
  \bibinfo {author} {\bibfnamefont {B.}~\bibnamefont {Langer}}, \bibinfo
  {author} {\bibfnamefont {C.}~\bibnamefont {Graf}}, \bibinfo {author}
  {\bibfnamefont {M.~I.}\ \bibnamefont {Stockman}}, \bibinfo {author}
  {\bibfnamefont {F.}~\bibnamefont {Krausz}}, \bibinfo {author} {\bibfnamefont
  {E.}~\bibnamefont {Ruhl}}, \ and\ \bibinfo {author} {\bibfnamefont {M.~F.}\
  \bibnamefont {Kling}},\ }\href {\doibase 10.1038/nphys1983} {\bibfield
  {journal} {\bibinfo  {journal} {Nat. Phys.}\ }\textbf {\bibinfo {volume}
  {7}},\ \bibinfo {pages} {656} (\bibinfo {year} {2011})}\BibitemShut {NoStop}%
\bibitem [{\citenamefont {Kr\"{u}ger}\ \emph {et~al.}(2011)\citenamefont
  {Kr\"{u}ger}, \citenamefont {Schenk},\ and\ \citenamefont
  {Hommelhoff}}]{KruSchHom2011}%
  \BibitemOpen
  \bibfield  {author} {\bibinfo {author} {\bibfnamefont {M.}~\bibnamefont
  {Kr\"{u}ger}}, \bibinfo {author} {\bibfnamefont {M.}~\bibnamefont {Schenk}},
  \ and\ \bibinfo {author} {\bibfnamefont {P.}~\bibnamefont {Hommelhoff}},\
  }\href {\doibase 10.1038/nature10196} {\bibfield  {journal} {\bibinfo
  {journal} {Nature}\ }\textbf {\bibinfo {volume} {475}},\ \bibinfo {pages}
  {78} (\bibinfo {year} {2011})}\BibitemShut {NoStop}%
\bibitem [{\citenamefont {Schiffrin}\ \emph {et~al.}(2013)\citenamefont
  {Schiffrin}, \citenamefont {Paasch-Colberg}, \citenamefont {Karpowicz},
  \citenamefont {Apalkov}, \citenamefont {Gerster}, \citenamefont {Muhlbrandt},
  \citenamefont {Korbman}, \citenamefont {Reichert}, \citenamefont {Schultze},
  \citenamefont {Holzner}, \citenamefont {Barth}, \citenamefont {Kienberger},
  \citenamefont {Ernstorfer}, \citenamefont {Yakovlev}, \citenamefont
  {Stockman},\ and\ \citenamefont {Krausz}}]{SchPaaKar2012}%
  \BibitemOpen
  \bibfield  {author} {\bibinfo {author} {\bibfnamefont {A.}~\bibnamefont
  {Schiffrin}}, \bibinfo {author} {\bibfnamefont {T.}~\bibnamefont
  {Paasch-Colberg}}, \bibinfo {author} {\bibfnamefont {N.}~\bibnamefont
  {Karpowicz}}, \bibinfo {author} {\bibfnamefont {V.}~\bibnamefont {Apalkov}},
  \bibinfo {author} {\bibfnamefont {D.}~\bibnamefont {Gerster}}, \bibinfo
  {author} {\bibfnamefont {S.}~\bibnamefont {Muhlbrandt}}, \bibinfo {author}
  {\bibfnamefont {M.}~\bibnamefont {Korbman}}, \bibinfo {author} {\bibfnamefont
  {J.}~\bibnamefont {Reichert}}, \bibinfo {author} {\bibfnamefont
  {M.}~\bibnamefont {Schultze}}, \bibinfo {author} {\bibfnamefont
  {S.}~\bibnamefont {Holzner}}, \bibinfo {author} {\bibfnamefont {J.~V.}\
  \bibnamefont {Barth}}, \bibinfo {author} {\bibfnamefont {R.}~\bibnamefont
  {Kienberger}}, \bibinfo {author} {\bibfnamefont {R.}~\bibnamefont
  {Ernstorfer}}, \bibinfo {author} {\bibfnamefont {V.~S.}\ \bibnamefont
  {Yakovlev}}, \bibinfo {author} {\bibfnamefont {M.~I.}\ \bibnamefont
  {Stockman}}, \ and\ \bibinfo {author} {\bibfnamefont {F.}~\bibnamefont
  {Krausz}},\ }\href {\doibase 10.1038/nature11567} {\bibfield  {journal}
  {\bibinfo  {journal} {Nature}\ }\textbf {\bibinfo {volume} {493}},\ \bibinfo
  {pages} {70} (\bibinfo {year} {2013})}\BibitemShut {NoStop}%
\bibitem [{\citenamefont {Wirth}\ \emph {et~al.}(2011)\citenamefont {Wirth},
  \citenamefont {Hassan}, \citenamefont {Grgura\v{s}}, \citenamefont {Gagnon},
  \citenamefont {Moulet}, \citenamefont {Luu}, \citenamefont {Pabst},
  \citenamefont {Santra}, \citenamefont {Alahmed}, \citenamefont {Azzeer},
  \citenamefont {Yakovlev}, \citenamefont {Pervak}, \citenamefont {Krausz},\
  and\ \citenamefont {Goulielmakis}}]{WirHasGrg2011}%
  \BibitemOpen
  \bibfield  {author} {\bibinfo {author} {\bibfnamefont {A.}~\bibnamefont
  {Wirth}}, \bibinfo {author} {\bibfnamefont {M.~T.}\ \bibnamefont {Hassan}},
  \bibinfo {author} {\bibfnamefont {I.}~\bibnamefont {Grgura\v{s}}}, \bibinfo
  {author} {\bibfnamefont {J.}~\bibnamefont {Gagnon}}, \bibinfo {author}
  {\bibfnamefont {A.}~\bibnamefont {Moulet}}, \bibinfo {author} {\bibfnamefont
  {T.~T.}\ \bibnamefont {Luu}}, \bibinfo {author} {\bibfnamefont
  {S.}~\bibnamefont {Pabst}}, \bibinfo {author} {\bibfnamefont
  {R.}~\bibnamefont {Santra}}, \bibinfo {author} {\bibfnamefont {Z.~A.}\
  \bibnamefont {Alahmed}}, \bibinfo {author} {\bibfnamefont {A.~M.}\
  \bibnamefont {Azzeer}}, \bibinfo {author} {\bibfnamefont {V.~S.}\
  \bibnamefont {Yakovlev}}, \bibinfo {author} {\bibfnamefont {V.}~\bibnamefont
  {Pervak}}, \bibinfo {author} {\bibfnamefont {F.}~\bibnamefont {Krausz}}, \
  and\ \bibinfo {author} {\bibfnamefont {E.}~\bibnamefont {Goulielmakis}},\
  }\href {\doibase 10.1126/science.1210268} {\bibfield  {journal} {\bibinfo
  {journal} {Science}\ }\textbf {\bibinfo {volume} {334}},\ \bibinfo {pages}
  {195} (\bibinfo {year} {2011})}\BibitemShut {NoStop}%
\bibitem [{\citenamefont {Stebbings}\ \emph {et~al.}(2011)\citenamefont
  {Stebbings}, \citenamefont {S\"{u}{\ss}mann}, \citenamefont {Yang},
  \citenamefont {Scrinzi}, \citenamefont {Durach}, \citenamefont {Rusina},
  \citenamefont {Stockman},\ and\ \citenamefont {Kling}}]{SteSueYan2011}%
  \BibitemOpen
  \bibfield  {author} {\bibinfo {author} {\bibfnamefont {S.~L.}\ \bibnamefont
  {Stebbings}}, \bibinfo {author} {\bibfnamefont {F.}~\bibnamefont
  {S\"{u}{\ss}mann}}, \bibinfo {author} {\bibfnamefont {Y.-Y.}\ \bibnamefont
  {Yang}}, \bibinfo {author} {\bibfnamefont {A.}~\bibnamefont {Scrinzi}},
  \bibinfo {author} {\bibfnamefont {M.}~\bibnamefont {Durach}}, \bibinfo
  {author} {\bibfnamefont {A.}~\bibnamefont {Rusina}}, \bibinfo {author}
  {\bibfnamefont {M.~I.}\ \bibnamefont {Stockman}}, \ and\ \bibinfo {author}
  {\bibfnamefont {M.~F.}\ \bibnamefont {Kling}},\ }\href {\doibase
  10.1088/1367-2630/13/7/073010} {\bibfield  {journal} {\bibinfo  {journal}
  {New J. Phys.}\ }\textbf {\bibinfo {volume} {13}},\ \bibinfo {pages} {073010}
  (\bibinfo {year} {2011})}\BibitemShut {NoStop}%
\bibitem [{\citenamefont {Roudnev}\ and\ \citenamefont
  {Esry}(2007)}]{RouEsr2007}%
  \BibitemOpen
  \bibfield  {author} {\bibinfo {author} {\bibfnamefont {V.}~\bibnamefont
  {Roudnev}}\ and\ \bibinfo {author} {\bibfnamefont {B.~D.}\ \bibnamefont
  {Esry}},\ }\href {\doibase 10.1103/physrevlett.99.220406} {\bibfield
  {journal} {\bibinfo  {journal} {Phys. Rev. Lett.}\ }\textbf {\bibinfo
  {volume} {99}},\ \bibinfo {pages} {220406} (\bibinfo {year}
  {2007})}\BibitemShut {NoStop}%
\bibitem [{\citenamefont {Kim}\ \emph {et~al.}(2008)\citenamefont {Kim},
  \citenamefont {Jin}, \citenamefont {Kim}, \citenamefont {Park}, \citenamefont
  {Kim},\ and\ \citenamefont {Kim}}]{KimJinKim2008}%
  \BibitemOpen
  \bibfield  {author} {\bibinfo {author} {\bibfnamefont {S.}~\bibnamefont
  {Kim}}, \bibinfo {author} {\bibfnamefont {J.}~\bibnamefont {Jin}}, \bibinfo
  {author} {\bibfnamefont {Y.-J.}\ \bibnamefont {Kim}}, \bibinfo {author}
  {\bibfnamefont {I.-Y.}\ \bibnamefont {Park}}, \bibinfo {author}
  {\bibfnamefont {Y.}~\bibnamefont {Kim}}, \ and\ \bibinfo {author}
  {\bibfnamefont {S.-W.}\ \bibnamefont {Kim}},\ }\href {\doibase
  10.1038/nature07012} {\bibfield  {journal} {\bibinfo  {journal} {Nature}\
  }\textbf {\bibinfo {volume} {453}},\ \bibinfo {pages} {757} (\bibinfo {year}
  {2008})}\BibitemShut {NoStop}%
\bibitem [{\citenamefont {Park}\ \emph {et~al.}(2011)\citenamefont {Park},
  \citenamefont {Kim}, \citenamefont {Choi}, \citenamefont {Lee}, \citenamefont
  {Kim}, \citenamefont {Kling}, \citenamefont {Stockman},\ and\ \citenamefont
  {Kim}}]{ParKimCho2011}%
  \BibitemOpen
  \bibfield  {author} {\bibinfo {author} {\bibfnamefont {I.-Y.}\ \bibnamefont
  {Park}}, \bibinfo {author} {\bibfnamefont {S.}~\bibnamefont {Kim}}, \bibinfo
  {author} {\bibfnamefont {J.}~\bibnamefont {Choi}}, \bibinfo {author}
  {\bibfnamefont {D.-H.}\ \bibnamefont {Lee}}, \bibinfo {author} {\bibfnamefont
  {Y.-J.}\ \bibnamefont {Kim}}, \bibinfo {author} {\bibfnamefont {M.~F.}\
  \bibnamefont {Kling}}, \bibinfo {author} {\bibfnamefont {M.~I.}\ \bibnamefont
  {Stockman}}, \ and\ \bibinfo {author} {\bibfnamefont {S.-W.}\ \bibnamefont
  {Kim}},\ }\href {\doibase 10.1038/nphoton.2011.258} {\bibfield  {journal}
  {\bibinfo  {journal} {Nat. Phot.}\ }\textbf {\bibinfo {volume} {5}},\
  \bibinfo {pages} {677} (\bibinfo {year} {2011})}\BibitemShut {NoStop}%
\bibitem [{\citenamefont {Husakou}\ \emph
  {et~al.}(2011{\natexlab{a}})\citenamefont {Husakou}, \citenamefont
  {Kelkensberg}, \citenamefont {Herrmann},\ and\ \citenamefont
  {Vrakking}}]{HusKelHer2011}%
  \BibitemOpen
  \bibfield  {author} {\bibinfo {author} {\bibfnamefont {A.}~\bibnamefont
  {Husakou}}, \bibinfo {author} {\bibfnamefont {F.}~\bibnamefont
  {Kelkensberg}}, \bibinfo {author} {\bibfnamefont {J.}~\bibnamefont
  {Herrmann}}, \ and\ \bibinfo {author} {\bibfnamefont {M.~J.~J.}\ \bibnamefont
  {Vrakking}},\ }\href {\doibase 10.1364/oe.19.025346} {\bibfield  {journal}
  {\bibinfo  {journal} {Opt. Express}\ }\textbf {\bibinfo {volume} {19}},\
  \bibinfo {pages} {25346} (\bibinfo {year} {2011}{\natexlab{a}})}\BibitemShut
  {NoStop}%
\bibitem [{\citenamefont {Ciappina}\ \emph {et~al.}(2012)\citenamefont
  {Ciappina}, \citenamefont {Biegert}, \citenamefont {Quidant},\ and\
  \citenamefont {Lewenstein}}]{CiaBieQui2012}%
  \BibitemOpen
  \bibfield  {author} {\bibinfo {author} {\bibfnamefont {M.~F.}\ \bibnamefont
  {Ciappina}}, \bibinfo {author} {\bibfnamefont {J.}~\bibnamefont {Biegert}},
  \bibinfo {author} {\bibfnamefont {R.}~\bibnamefont {Quidant}}, \ and\
  \bibinfo {author} {\bibfnamefont {M.}~\bibnamefont {Lewenstein}},\ }\href
  {\doibase 10.1103/physreva.85.033828} {\bibfield  {journal} {\bibinfo
  {journal} {Phys. Rev. A}\ }\textbf {\bibinfo {volume} {85}},\ \bibinfo
  {pages} {033828} (\bibinfo {year} {2012})}\BibitemShut {NoStop}%
\bibitem [{\citenamefont {Sivis}\ \emph {et~al.}(2012)\citenamefont {Sivis},
  \citenamefont {Duwe}, \citenamefont {Abel},\ and\ \citenamefont
  {Ropers}}]{SivDuwAbe2012}%
  \BibitemOpen
  \bibfield  {author} {\bibinfo {author} {\bibfnamefont {M.}~\bibnamefont
  {Sivis}}, \bibinfo {author} {\bibfnamefont {M.}~\bibnamefont {Duwe}},
  \bibinfo {author} {\bibfnamefont {B.}~\bibnamefont {Abel}}, \ and\ \bibinfo
  {author} {\bibfnamefont {C.}~\bibnamefont {Ropers}},\ }\href {\doibase
  10.1038/nature10978} {\bibfield  {journal} {\bibinfo  {journal} {Nature}\
  }\textbf {\bibinfo {volume} {485}},\ \bibinfo {pages} {E1} (\bibinfo {year}
  {2012})}\BibitemShut {NoStop}%
\bibitem [{\citenamefont {Park}\ \emph {et~al.}(2013)\citenamefont {Park},
  \citenamefont {Choi}, \citenamefont {Lee}, \citenamefont {Han}, \citenamefont
  {Kim},\ and\ \citenamefont {Kim}}]{ParChoLee2013}%
  \BibitemOpen
  \bibfield  {author} {\bibinfo {author} {\bibfnamefont {I.-Y.}\ \bibnamefont
  {Park}}, \bibinfo {author} {\bibfnamefont {J.}~\bibnamefont {Choi}}, \bibinfo
  {author} {\bibfnamefont {D.-H.}\ \bibnamefont {Lee}}, \bibinfo {author}
  {\bibfnamefont {S.}~\bibnamefont {Han}}, \bibinfo {author} {\bibfnamefont
  {S.}~\bibnamefont {Kim}}, \ and\ \bibinfo {author} {\bibfnamefont {S.-W.}\
  \bibnamefont {Kim}},\ }\href {\doibase 10.1002/andp.201200160} {\bibfield
  {journal} {\bibinfo  {journal} {Ann. Phys.}\ }\textbf {\bibinfo {volume}
  {525}},\ \bibinfo {pages} {87} (\bibinfo {year} {2013})}\BibitemShut
  {NoStop}%
\bibitem [{\citenamefont {Corkum}\ \emph {et~al.}(1994)\citenamefont {Corkum},
  \citenamefont {Burnett},\ and\ \citenamefont {Ivanov}}]{CorBurIva1994}%
  \BibitemOpen
  \bibfield  {author} {\bibinfo {author} {\bibfnamefont {P.~B.}\ \bibnamefont
  {Corkum}}, \bibinfo {author} {\bibfnamefont {N.~H.}\ \bibnamefont {Burnett}},
  \ and\ \bibinfo {author} {\bibfnamefont {M.~Y.}\ \bibnamefont {Ivanov}},\
  }\href {\doibase 10.1364/ol.19.001870} {\bibfield  {journal} {\bibinfo
  {journal} {Opt. Lett.}\ }\textbf {\bibinfo {volume} {19}},\ \bibinfo {pages}
  {1870} (\bibinfo {year} {1994})}\BibitemShut {NoStop}%
\bibitem [{\citenamefont {Shan}\ \emph {et~al.}(2005)\citenamefont {Shan},
  \citenamefont {Ghimire},\ and\ \citenamefont {Chang}}]{ShaGhiCha2005}%
  \BibitemOpen
  \bibfield  {author} {\bibinfo {author} {\bibfnamefont {B.}~\bibnamefont
  {Shan}}, \bibinfo {author} {\bibfnamefont {S.}~\bibnamefont {Ghimire}}, \
  and\ \bibinfo {author} {\bibfnamefont {Z.}~\bibnamefont {Chang}},\ }\href
  {\doibase 10.1080/09500340410001729573} {\bibfield  {journal} {\bibinfo
  {journal} {J. Mod. Opt.}\ }\textbf {\bibinfo {volume} {52}},\ \bibinfo
  {pages} {277} (\bibinfo {year} {2005})}\BibitemShut {NoStop}%
\bibitem [{\citenamefont {Sansone}\ \emph {et~al.}(2006)\citenamefont
  {Sansone}, \citenamefont {Benedetti}, \citenamefont {Calegari}, \citenamefont
  {Vozzi}, \citenamefont {Avaldi}, \citenamefont {Flammini}, \citenamefont
  {Poletto}, \citenamefont {Villoresi}, \citenamefont {Altucci}, \citenamefont
  {Velotta}, \citenamefont {Stagira}, \citenamefont {De~Silvestri},\ and\
  \citenamefont {Nisoli}}]{SanBenCal2006}%
  \BibitemOpen
  \bibfield  {author} {\bibinfo {author} {\bibfnamefont {G.}~\bibnamefont
  {Sansone}}, \bibinfo {author} {\bibfnamefont {E.}~\bibnamefont {Benedetti}},
  \bibinfo {author} {\bibfnamefont {F.}~\bibnamefont {Calegari}}, \bibinfo
  {author} {\bibfnamefont {C.}~\bibnamefont {Vozzi}}, \bibinfo {author}
  {\bibfnamefont {L.}~\bibnamefont {Avaldi}}, \bibinfo {author} {\bibfnamefont
  {R.}~\bibnamefont {Flammini}}, \bibinfo {author} {\bibfnamefont
  {L.}~\bibnamefont {Poletto}}, \bibinfo {author} {\bibfnamefont
  {P.}~\bibnamefont {Villoresi}}, \bibinfo {author} {\bibfnamefont
  {C.}~\bibnamefont {Altucci}}, \bibinfo {author} {\bibfnamefont
  {R.}~\bibnamefont {Velotta}}, \bibinfo {author} {\bibfnamefont
  {S.}~\bibnamefont {Stagira}}, \bibinfo {author} {\bibfnamefont
  {S.}~\bibnamefont {De~Silvestri}}, \ and\ \bibinfo {author} {\bibfnamefont
  {M.}~\bibnamefont {Nisoli}},\ }\href {\doibase 10.1126/science.1132838}
  {\bibfield  {journal} {\bibinfo  {journal} {Science}\ }\textbf {\bibinfo
  {volume} {314}},\ \bibinfo {pages} {443} (\bibinfo {year}
  {2006})}\BibitemShut {NoStop}%
\bibitem [{\citenamefont {Christov}\ \emph {et~al.}(1997)\citenamefont
  {Christov}, \citenamefont {Murnane},\ and\ \citenamefont
  {Kapteyn}}]{Christov97}%
  \BibitemOpen
  \bibfield  {author} {\bibinfo {author} {\bibfnamefont {I.~P.}\ \bibnamefont
  {Christov}}, \bibinfo {author} {\bibfnamefont {M.~M.}\ \bibnamefont
  {Murnane}}, \ and\ \bibinfo {author} {\bibfnamefont {H.~C.}\ \bibnamefont
  {Kapteyn}},\ }\href {\doibase 10.1103/PhysRevLett.78.1251} {\bibfield
  {journal} {\bibinfo  {journal} {Phys. Rev. Lett.}\ }\textbf {\bibinfo
  {volume} {78}},\ \bibinfo {pages} {1251} (\bibinfo {year}
  {1997})}\BibitemShut {NoStop}%
\bibitem [{\citenamefont {Brabec}\ and\ \citenamefont
  {Krausz}(2000)}]{BraKra2000}%
  \BibitemOpen
  \bibfield  {author} {\bibinfo {author} {\bibfnamefont {T.}~\bibnamefont
  {Brabec}}\ and\ \bibinfo {author} {\bibfnamefont {F.}~\bibnamefont
  {Krausz}},\ }\href {\doibase 10.1103/revmodphys.72.545} {\bibfield  {journal}
  {\bibinfo  {journal} {Rev. Mod. Phys}\ }\textbf {\bibinfo {volume} {72}},\
  \bibinfo {pages} {545} (\bibinfo {year} {2000})}\BibitemShut {NoStop}%
\bibitem [{\citenamefont {Mashiko}\ \emph {et~al.}(2008)\citenamefont
  {Mashiko}, \citenamefont {Gilbertson}, \citenamefont {Li}, \citenamefont
  {Khan}, \citenamefont {Shakya}, \citenamefont {Moon},\ and\ \citenamefont
  {Chang}}]{MasGilLi2008}%
  \BibitemOpen
  \bibfield  {author} {\bibinfo {author} {\bibfnamefont {H.}~\bibnamefont
  {Mashiko}}, \bibinfo {author} {\bibfnamefont {S.}~\bibnamefont {Gilbertson}},
  \bibinfo {author} {\bibfnamefont {C.}~\bibnamefont {Li}}, \bibinfo {author}
  {\bibfnamefont {S.~D.}\ \bibnamefont {Khan}}, \bibinfo {author}
  {\bibfnamefont {M.~M.}\ \bibnamefont {Shakya}}, \bibinfo {author}
  {\bibfnamefont {E.}~\bibnamefont {Moon}}, \ and\ \bibinfo {author}
  {\bibfnamefont {Z.}~\bibnamefont {Chang}},\ }\href {\doibase
  10.1103/physrevlett.100.103906} {\bibfield  {journal} {\bibinfo  {journal}
  {Phys. Rev. Lett.}\ }\textbf {\bibinfo {volume} {100}},\ \bibinfo {pages}
  {103906} (\bibinfo {year} {2008})}\BibitemShut {NoStop}%
\bibitem [{\citenamefont {Feng}\ \emph {et~al.}(2009)\citenamefont {Feng},
  \citenamefont {Gilbertson}, \citenamefont {Mashiko}, \citenamefont {Wang},
  \citenamefont {Khan}, \citenamefont {Chini}, \citenamefont {Wu},
  \citenamefont {Zhao},\ and\ \citenamefont {Chang}}]{FenGilMas2009}%
  \BibitemOpen
  \bibfield  {author} {\bibinfo {author} {\bibfnamefont {X.}~\bibnamefont
  {Feng}}, \bibinfo {author} {\bibfnamefont {S.}~\bibnamefont {Gilbertson}},
  \bibinfo {author} {\bibfnamefont {H.}~\bibnamefont {Mashiko}}, \bibinfo
  {author} {\bibfnamefont {H.}~\bibnamefont {Wang}}, \bibinfo {author}
  {\bibfnamefont {S.~D.}\ \bibnamefont {Khan}}, \bibinfo {author}
  {\bibfnamefont {M.}~\bibnamefont {Chini}}, \bibinfo {author} {\bibfnamefont
  {Y.}~\bibnamefont {Wu}}, \bibinfo {author} {\bibfnamefont {K.}~\bibnamefont
  {Zhao}}, \ and\ \bibinfo {author} {\bibfnamefont {Z.}~\bibnamefont {Chang}},\
  }\href {\doibase 10.1103/physrevlett.103.183901} {\bibfield  {journal}
  {\bibinfo  {journal} {Phys. Rev. Lett.}\ }\textbf {\bibinfo {volume} {103}},\
  \bibinfo {pages} {183901} (\bibinfo {year} {2009})}\BibitemShut {NoStop}%
\bibitem [{\citenamefont {Haynes}(2013)}]{CRCHandbook}%
  \BibitemOpen
  \bibinfo {editor} {\bibfnamefont {W.~M.}\ \bibnamefont {Haynes}},\ ed.,\
  \href {http://www.hbcpnetbase.com/} {\emph {\bibinfo {title} {{CRC} Handbook
  of Chemistry and Physics}}},\ \bibinfo {edition} {93rd}\ ed.\ (\bibinfo
  {publisher} {CRC Press/Taylor and Francis},\ \bibinfo {address} {Boca Raton,
  FL},\ \bibinfo {year} {2013})\ \bibinfo {note} {internet Version
  2013}\BibitemShut {NoStop}%
\bibitem [{\citenamefont {McMahon}\ \emph {et~al.}(2009)\citenamefont
  {McMahon}, \citenamefont {Gray},\ and\ \citenamefont
  {Schatz}}]{McmGraSch2009}%
  \BibitemOpen
  \bibfield  {author} {\bibinfo {author} {\bibfnamefont {J.~M.}\ \bibnamefont
  {McMahon}}, \bibinfo {author} {\bibfnamefont {S.~K.}\ \bibnamefont {Gray}}, \
  and\ \bibinfo {author} {\bibfnamefont {G.~C.}\ \bibnamefont {Schatz}},\
  }\href {\doibase 10.1103/physrevlett.103.097403} {\bibfield  {journal}
  {\bibinfo  {journal} {Phys. Rev. Lett.}\ }\textbf {\bibinfo {volume} {103}},\
  \bibinfo {pages} {097403} (\bibinfo {year} {2009})}\BibitemShut {NoStop}%
\bibitem [{\citenamefont {McMahon}\ \emph {et~al.}(2010)\citenamefont
  {McMahon}, \citenamefont {Gray},\ and\ \citenamefont
  {Schatz}}]{McMGraSch2010}%
  \BibitemOpen
  \bibfield  {author} {\bibinfo {author} {\bibfnamefont {J.~M.}\ \bibnamefont
  {McMahon}}, \bibinfo {author} {\bibfnamefont {S.~K.}\ \bibnamefont {Gray}}, \
  and\ \bibinfo {author} {\bibfnamefont {G.~C.}\ \bibnamefont {Schatz}},\
  }\href {\doibase 10.1103/physrevb.82.035423} {\bibfield  {journal} {\bibinfo
  {journal} {Phys. Rev. B}\ }\textbf {\bibinfo {volume} {82}},\ \bibinfo
  {pages} {035423} (\bibinfo {year} {2010})}\BibitemShut {NoStop}%
\bibitem [{\citenamefont {David}\ and\ \citenamefont {Garc\'{\i}a~de
  Abajo}(2011)}]{DavGar2011}%
  \BibitemOpen
  \bibfield  {author} {\bibinfo {author} {\bibfnamefont {C.}~\bibnamefont
  {David}}\ and\ \bibinfo {author} {\bibfnamefont {F.~J.}\ \bibnamefont
  {Garc\'{\i}a~de Abajo}},\ }\href {\doibase 10.1021/jp204261u} {\bibfield
  {journal} {\bibinfo  {journal} {J. Phys. Chem. C}\ }\textbf {\bibinfo
  {volume} {115}},\ \bibinfo {pages} {19470} (\bibinfo {year}
  {2011})}\BibitemShut {NoStop}%
\bibitem [{scu(2012)}]{scuff-em}%
  \BibitemOpen
  \href {http://homerreid.com/scuff-EM} {\enquote {\bibinfo {title}
  {\textsc{scuff-em}},}\ } (\bibinfo {year} {2012})\BibitemShut {NoStop}%
\bibitem [{\citenamefont {Reid}\ \emph {et~al.}(2009)\citenamefont {Reid},
  \citenamefont {Rodriguez}, \citenamefont {White},\ and\ \citenamefont
  {Johnson}}]{ReiRodWhi2009}%
  \BibitemOpen
  \bibfield  {author} {\bibinfo {author} {\bibfnamefont {M.~T.~H.}\
  \bibnamefont {Reid}}, \bibinfo {author} {\bibfnamefont {A.~W.}\ \bibnamefont
  {Rodriguez}}, \bibinfo {author} {\bibfnamefont {J.}~\bibnamefont {White}}, \
  and\ \bibinfo {author} {\bibfnamefont {S.~G.}\ \bibnamefont {Johnson}},\
  }\href {\doibase 10.1103/physrevlett.103.040401} {\bibfield  {journal}
  {\bibinfo  {journal} {Phys. Rev. Lett.}\ }\textbf {\bibinfo {volume} {103}},\
  \bibinfo {pages} {040401} (\bibinfo {year} {2009})}\BibitemShut {NoStop}%
\bibitem [{\citenamefont {Harrington}(1993)}]{Har1993}%
  \BibitemOpen
  \bibfield  {author} {\bibinfo {author} {\bibfnamefont {R.~F.}\ \bibnamefont
  {Harrington}},\ }\href {http://www.worldcat.org/isbn/9780198592174} {\emph
  {\bibinfo {title} {Field computation by moment methods}}}\ (\bibinfo
  {publisher} {IEEE Press},\ \bibinfo {address} {Piscataway, NJ},\ \bibinfo
  {year} {1993})\BibitemShut {NoStop}%
\bibitem [{\citenamefont {Scrinzi}\ \emph {et~al.}(2006)\citenamefont
  {Scrinzi}, \citenamefont {Ivanov}, \citenamefont {Kienberger},\ and\
  \citenamefont {Villeneuve}}]{ScrIvaKie2006}%
  \BibitemOpen
  \bibfield  {author} {\bibinfo {author} {\bibfnamefont {A.}~\bibnamefont
  {Scrinzi}}, \bibinfo {author} {\bibfnamefont {M.~Y.}\ \bibnamefont {Ivanov}},
  \bibinfo {author} {\bibfnamefont {R.}~\bibnamefont {Kienberger}}, \ and\
  \bibinfo {author} {\bibfnamefont {D.~M.}\ \bibnamefont {Villeneuve}},\ }\href
  {\doibase 10.1088/0953-4075/39/1/R01} {\bibfield  {journal} {\bibinfo
  {journal} {J. Phys. B}\ }\textbf {\bibinfo {volume} {39}},\ \bibinfo {pages}
  {R1} (\bibinfo {year} {2006})}\BibitemShut {NoStop}%
\bibitem [{\citenamefont {Corkum}\ and\ \citenamefont
  {Krausz}(2007)}]{Corkum07}%
  \BibitemOpen
  \bibfield  {author} {\bibinfo {author} {\bibfnamefont {P.~B.}\ \bibnamefont
  {Corkum}}\ and\ \bibinfo {author} {\bibfnamefont {F.}~\bibnamefont
  {Krausz}},\ }\href {\doibase 10.1038/nphys620} {\bibfield  {journal}
  {\bibinfo  {journal} {Nat. Phys.}\ }\textbf {\bibinfo {volume} {3}},\
  \bibinfo {pages} {381} (\bibinfo {year} {2007})}\BibitemShut {NoStop}%
\bibitem [{\citenamefont {Baumert}\ \emph {et~al.}(1997)\citenamefont
  {Baumert}, \citenamefont {Brixner}, \citenamefont {Seyfried}, \citenamefont
  {Strehle},\ and\ \citenamefont {Gerber}}]{BauBriSey1997}%
  \BibitemOpen
  \bibfield  {author} {\bibinfo {author} {\bibfnamefont {T.}~\bibnamefont
  {Baumert}}, \bibinfo {author} {\bibfnamefont {T.}~\bibnamefont {Brixner}},
  \bibinfo {author} {\bibfnamefont {V.}~\bibnamefont {Seyfried}}, \bibinfo
  {author} {\bibfnamefont {M.}~\bibnamefont {Strehle}}, \ and\ \bibinfo
  {author} {\bibfnamefont {G.}~\bibnamefont {Gerber}},\ }\href {\doibase
  10.1007/s003400050346} {\bibfield  {journal} {\bibinfo  {journal} {Appl.
  Phys. B}\ }\textbf {\bibinfo {volume} {65}},\ \bibinfo {pages} {779}
  (\bibinfo {year} {1997})}\BibitemShut {NoStop}%
\bibitem [{\citenamefont {Weiner}(2000)}]{Wei2000}%
  \BibitemOpen
  \bibfield  {author} {\bibinfo {author} {\bibfnamefont {A.~M.}\ \bibnamefont
  {Weiner}},\ }\href {\doibase 10.1063/1.1150614} {\bibfield  {journal}
  {\bibinfo  {journal} {Review of Scientific Instruments}\ }\textbf {\bibinfo
  {volume} {71}},\ \bibinfo {pages} {1929} (\bibinfo {year}
  {2000})}\BibitemShut {NoStop}%
\bibitem [{\citenamefont {Lewenstein}\ \emph {et~al.}(1994)\citenamefont
  {Lewenstein}, \citenamefont {Balcou}, \citenamefont {Ivanov}, \citenamefont
  {L'Huillier},\ and\ \citenamefont {Corkum}}]{Lewenstein94}%
  \BibitemOpen
  \bibfield  {author} {\bibinfo {author} {\bibfnamefont {M.}~\bibnamefont
  {Lewenstein}}, \bibinfo {author} {\bibfnamefont {P.}~\bibnamefont {Balcou}},
  \bibinfo {author} {\bibfnamefont {M.~Y.}\ \bibnamefont {Ivanov}}, \bibinfo
  {author} {\bibfnamefont {A.}~\bibnamefont {L'Huillier}}, \ and\ \bibinfo
  {author} {\bibfnamefont {P.~B.}\ \bibnamefont {Corkum}},\ }\href {\doibase
  10.1103/PhysRevA.49.2117} {\bibfield  {journal} {\bibinfo  {journal} {Phys.
  Rev. A}\ }\textbf {\bibinfo {volume} {49}},\ \bibinfo {pages} {2117}
  (\bibinfo {year} {1994})}\BibitemShut {NoStop}%
\bibitem [{\citenamefont {Ivanov}\ \emph {et~al.}(1996)\citenamefont {Ivanov},
  \citenamefont {Brabec},\ and\ \citenamefont {Burnett}}]{IvaBraBur1996}%
  \BibitemOpen
  \bibfield  {author} {\bibinfo {author} {\bibfnamefont {M.~Y.}\ \bibnamefont
  {Ivanov}}, \bibinfo {author} {\bibfnamefont {T.}~\bibnamefont {Brabec}}, \
  and\ \bibinfo {author} {\bibfnamefont {N.}~\bibnamefont {Burnett}},\ }\href
  {\doibase 10.1103/physreva.54.742} {\bibfield  {journal} {\bibinfo  {journal}
  {Phys. Rev. A}\ }\textbf {\bibinfo {volume} {54}},\ \bibinfo {pages} {742}
  (\bibinfo {year} {1996})}\BibitemShut {NoStop}%
\bibitem [{\citenamefont {Husakou}\ \emph
  {et~al.}(2011{\natexlab{b}})\citenamefont {Husakou}, \citenamefont {Im},\
  and\ \citenamefont {Herrmann}}]{HusImHer2011}%
  \BibitemOpen
  \bibfield  {author} {\bibinfo {author} {\bibfnamefont {A.}~\bibnamefont
  {Husakou}}, \bibinfo {author} {\bibfnamefont {S.~J.}\ \bibnamefont {Im}}, \
  and\ \bibinfo {author} {\bibfnamefont {J.}~\bibnamefont {Herrmann}},\ }\href
  {\doibase 10.1103/physreva.83.043839} {\bibfield  {journal} {\bibinfo
  {journal} {Phys. Rev. A}\ }\textbf {\bibinfo {volume} {83}},\ \bibinfo
  {pages} {043839} (\bibinfo {year} {2011}{\natexlab{b}})}\BibitemShut
  {NoStop}%
\bibitem [{\citenamefont {Constant}\ \emph {et~al.}(1997)\citenamefont
  {Constant}, \citenamefont {Taranukhin}, \citenamefont {Stolow},\ and\
  \citenamefont {Corkum}}]{ConTarSto1997}%
  \BibitemOpen
  \bibfield  {author} {\bibinfo {author} {\bibfnamefont {E.}~\bibnamefont
  {Constant}}, \bibinfo {author} {\bibfnamefont {V.~D.}\ \bibnamefont
  {Taranukhin}}, \bibinfo {author} {\bibfnamefont {A.}~\bibnamefont {Stolow}},
  \ and\ \bibinfo {author} {\bibfnamefont {P.~B.}\ \bibnamefont {Corkum}},\
  }\href {\doibase 10.1103/physreva.56.3870} {\bibfield  {journal} {\bibinfo
  {journal} {Phys. Rev. A}\ }\textbf {\bibinfo {volume} {56}},\ \bibinfo
  {pages} {3870} (\bibinfo {year} {1997})}\BibitemShut {NoStop}%
\bibitem [{\citenamefont {Goulielmakis}\ \emph {et~al.}(2004)\citenamefont
  {Goulielmakis}, \citenamefont {Uiberacker}, \citenamefont {Kienberger},
  \citenamefont {Baltuska}, \citenamefont {Yakovlev}, \citenamefont {Scrinzi},
  \citenamefont {Westerwalbesloh}, \citenamefont {Kleineberg}, \citenamefont
  {Heinzmann}, \citenamefont {Drescher},\ and\ \citenamefont
  {Krausz}}]{GouUibKie2004}%
  \BibitemOpen
  \bibfield  {author} {\bibinfo {author} {\bibfnamefont {E.}~\bibnamefont
  {Goulielmakis}}, \bibinfo {author} {\bibfnamefont {M.}~\bibnamefont
  {Uiberacker}}, \bibinfo {author} {\bibfnamefont {R.}~\bibnamefont
  {Kienberger}}, \bibinfo {author} {\bibfnamefont {A.}~\bibnamefont
  {Baltuska}}, \bibinfo {author} {\bibfnamefont {V.}~\bibnamefont {Yakovlev}},
  \bibinfo {author} {\bibfnamefont {A.}~\bibnamefont {Scrinzi}}, \bibinfo
  {author} {\bibfnamefont {T.}~\bibnamefont {Westerwalbesloh}}, \bibinfo
  {author} {\bibfnamefont {U.}~\bibnamefont {Kleineberg}}, \bibinfo {author}
  {\bibfnamefont {U.}~\bibnamefont {Heinzmann}}, \bibinfo {author}
  {\bibfnamefont {M.}~\bibnamefont {Drescher}}, \ and\ \bibinfo {author}
  {\bibfnamefont {F.}~\bibnamefont {Krausz}},\ }\href {\doibase
  10.1126/science.1100866} {\bibfield  {journal} {\bibinfo  {journal}
  {Science}\ }\textbf {\bibinfo {volume} {305}},\ \bibinfo {pages} {1267}
  (\bibinfo {year} {2004})}\BibitemShut {NoStop}%
\bibitem [{\citenamefont {Raz}\ \emph {et~al.}(2011)\citenamefont {Raz},
  \citenamefont {Schwartz}, \citenamefont {Austin}, \citenamefont {Wyatt},
  \citenamefont {Schiavi}, \citenamefont {Smirnova}, \citenamefont {Nadler},
  \citenamefont {Walmsley}, \citenamefont {Oron},\ and\ \citenamefont
  {Dudovich}}]{RazSchAus2011}%
  \BibitemOpen
  \bibfield  {author} {\bibinfo {author} {\bibfnamefont {O.}~\bibnamefont
  {Raz}}, \bibinfo {author} {\bibfnamefont {O.}~\bibnamefont {Schwartz}},
  \bibinfo {author} {\bibfnamefont {D.}~\bibnamefont {Austin}}, \bibinfo
  {author} {\bibfnamefont {A.~S.}\ \bibnamefont {Wyatt}}, \bibinfo {author}
  {\bibfnamefont {A.}~\bibnamefont {Schiavi}}, \bibinfo {author} {\bibfnamefont
  {O.}~\bibnamefont {Smirnova}}, \bibinfo {author} {\bibfnamefont
  {B.}~\bibnamefont {Nadler}}, \bibinfo {author} {\bibfnamefont {I.~A.}\
  \bibnamefont {Walmsley}}, \bibinfo {author} {\bibfnamefont {D.}~\bibnamefont
  {Oron}}, \ and\ \bibinfo {author} {\bibfnamefont {N.}~\bibnamefont
  {Dudovich}},\ }\href {\doibase 10.1103/physrevlett.107.133902} {\bibfield
  {journal} {\bibinfo  {journal} {Phys. Rev. Lett.}\ }\textbf {\bibinfo
  {volume} {107}},\ \bibinfo {pages} {133902} (\bibinfo {year}
  {2011})}\BibitemShut {NoStop}%
\bibitem [{\citenamefont {Kim}\ \emph {et~al.}(2013)\citenamefont {Kim},
  \citenamefont {Zhang}, \citenamefont {Shiner}, \citenamefont {Kirkwood},
  \citenamefont {Frumker}, \citenamefont {Gariepy}, \citenamefont {Naumov},
  \citenamefont {Villeneuve},\ and\ \citenamefont {Corkum}}]{KimZhaShi2013}%
  \BibitemOpen
  \bibfield  {author} {\bibinfo {author} {\bibfnamefont {K.~T.}\ \bibnamefont
  {Kim}}, \bibinfo {author} {\bibfnamefont {C.}~\bibnamefont {Zhang}}, \bibinfo
  {author} {\bibfnamefont {A.~D.}\ \bibnamefont {Shiner}}, \bibinfo {author}
  {\bibfnamefont {S.~E.}\ \bibnamefont {Kirkwood}}, \bibinfo {author}
  {\bibfnamefont {E.}~\bibnamefont {Frumker}}, \bibinfo {author} {\bibfnamefont
  {G.}~\bibnamefont {Gariepy}}, \bibinfo {author} {\bibfnamefont
  {A.}~\bibnamefont {Naumov}}, \bibinfo {author} {\bibfnamefont {D.~M.}\
  \bibnamefont {Villeneuve}}, \ and\ \bibinfo {author} {\bibfnamefont {P.~B.}\
  \bibnamefont {Corkum}},\ }\href {\doibase 10.1038/nphys2525} {\bibfield
  {journal} {\bibinfo  {journal} {Nat. Phys.}\ }\textbf {\bibinfo {volume}
  {9}},\ \bibinfo {pages} {159} (\bibinfo {year} {2013})}\BibitemShut {NoStop}%
\bibitem [{\citenamefont {Rowan}(1990)}]{SUBPLEX}%
  \BibitemOpen
  \bibfield  {author} {\bibinfo {author} {\bibfnamefont {T.}~\bibnamefont
  {Rowan}},\ }\emph {\bibinfo {title} {Functional Stability Analysis of
  Numerical Algorithms}},\ \href@noop {} {Ph.D. thesis},\ \bibinfo  {school}
  {Department of Computer Sciences, University of Texas at Austin} (\bibinfo
  {year} {1990})\BibitemShut {NoStop}%
\bibitem [{\citenamefont {Johnson}(2012)}]{NLOPT}%
  \BibitemOpen
  \bibfield  {author} {\bibinfo {author} {\bibfnamefont {S.~G.}\ \bibnamefont
  {Johnson}},\ }\href {http://ab-initio.mit.edu/nlopt} {\enquote {\bibinfo
  {title} {The {NLopt} nonlinear-optimization package},}\ } (\bibinfo {year}
  {2012})\BibitemShut {NoStop}%
\end{thebibliography}
\end{document}